\documentclass[a4paper,11pt]{article}

\usepackage[utf8]{inputenc}
\usepackage{a4wide}
\usepackage{graphicx}
\graphicspath{{figures/}} 
\usepackage[small,bf]{caption}
\usepackage{amssymb}
\usepackage{amsmath}
\usepackage{mathrsfs}
\usepackage{slashed}
\usepackage{mathtools}
\usepackage{cite} 
\usepackage{color,xcolor}
\usepackage{comment}
\def\linkcolor{cyan!70!black}

\usepackage[
colorlinks=true
,urlcolor=\linkcolor
,anchorcolor=\linkcolor
,citecolor=\linkcolor
,filecolor=\linkcolor
,linkcolor=\linkcolor
,menucolor=\linkcolor
,linktocpage=true
,pdfproducer=medialab
,pdfa=true
]{hyperref}

\newcommand{\fref}[1]{Figure~\ref{#1}} 
\newcommand{\eref}[1]{Eq.\,(\ref{#1})}

\newcommand{\aref}[1]{Appendix~\ref{#1}}
\newcommand{\sref}[1]{Section~\ref{#1}}

\numberwithin{equation}{section}
\makeatletter
\g@addto@macro\bfseries{\boldmath}
\makeatother

\begin{document}

\begin{titlepage}
\begin{flushright}
IFT-UAM/CSIC-23-45
\end{flushright}
\vspace*{0.8cm}

\begin{center}
{\LARGE\bf Type-II Majoron Dark Matter}\\[0.8cm]
{\large 
Carla~Biggio$\,^a$,
Lorenzo~Calibbi$\,^b$,
Toshihiko~Ota$\,^{c,d}$,
and
Samuele~Zanchini$\,^a$\\[1.cm]
}
$^{a}$\,{\it Dipartimento di Fisica, Universit\`a di Genova and INFN -\\ 
Sezione di Genova, via Dodecaneso 33, 16146 Genova, Italy} \\[0.2cm]
$^{b}$\,{\it School of Physics, Nankai University, Tianjin 300071, China} \\[0.2cm]
$^{c}$\,{\it Departamento de F\'isica Te\'orica and Instituto de F\'{\i}sica Te\'orica, IFT-UAM/CSIC,\\
Universidad Aut\'onoma de Madrid, Cantoblanco, 28049, Madrid, Spain} 
\\[0.2cm]
$^{d}$\,{\it Departamento de F\'{i}sica, Facultad de Ciencias, Universidad de La Serena, Avenida Cisternas
1200, La Serena, Chile} 
\\
\end{center}
\vspace{0.8cm}

\begin{abstract}
\noindent 
We discuss in detail the possibility that the ``type-II majoron''\,\,---\,\,that is, the pseudo Nambu-Goldstone boson that arises in the context of the type-II seesaw mechanism if the lepton number is spontaneously broken by an additional singlet scalar\,\,---\,\,account for the dark matter (DM) observed in the universe.
We study the requirements the model's parameters have to fulfill in order to reproduce the measured DM relic abundance through two possible production mechanisms in the early universe, freeze-in and misalignment, both during a standard radiation-dominated era and early matter domination. We then study possible signals of type-II majoron DM and the present and expected constraints on the parameter space that can be obtained from cosmological observations, direct detection experiments, and present and future searches for decaying DM at neutrino telescopes and cosmic-ray experiments. We find that\,\,---\,\,depending on the majoron mass, the production mechanism, and the vacuum expectation value of the type-II triplet\,\,---\,\,all of the three decay modes (photons, electrons, neutrinos) of majoron DM particles can yield observable signals at future indirect searches for DM. Furthermore, in a corner of the parameter space,  detection of majoron DM is possible through electron recoil at running and future direct detection experiments.
\end{abstract}

\end{titlepage}

\tableofcontents

\section{Introduction}
\label{sec:introduction}

After a decade of operations, the Large Hadron Collider~(LHC) has given no conclusive evidence 
of new phenomena beyond those predicted by the Standard Model~(SM) of particle physics. In particular, no sign of new particles at the TeV scale so far has been found. 
Similarly, Dark Matter~(DM) direct detection experiments, that is, searches for nuclear recoils due to collisions with DM particles,
have been only providing ever more stringent constraints on models where DM is made of Weakly Interacting Massive Particles (WIMPs). In combination with the LHC bounds, this significantly reduced the viable parameter space of typical WIMP models,
see, \emph{e.g.}, Refs.~\cite{Arcadi:2017kky,Bottaro:2021snn}. 
While the WIMP paradigm is far from being excluded, these experimental facts have recently motivated an increased interest in alternative DM candidates, such as very light, feebly-coupled\,\,---\,\,thus more elusive\,\,---\,\,particles, see, \emph{e.g.}, Refs.\cite{Adams:2022pbo,Irastorza:2018dyq,Bernal:2017kxu,Arias:2012az} and the references therein.

Paradigmatic examples of such light DM candidates are provided by the QCD axion~\cite{Peccei:1977hh,Wilczek:1977pj,Weinberg:1977ma} and, more in general, axion-like particles (ALPs), that is, pseudo Nambu-Goldstone bosons (pNGBs) of global $U(1)$ symmetries spontaneously broken at high energies. 
These fields can thus be interpreted as light remnants of some high-energy new physics scale, usually related to outstanding problems of the SM, such as the strong CP problem~\cite{Peccei:1977hh,Wilczek:1977pj,Weinberg:1977ma}, the flavour puzzle~\cite{Davidson:1981zd,Wilczek:1982rv,Reiss:1982sq,Davidson:1983fy,Chang:1987hz,Berezhiani:1990jj,Berezhiani:1990wn,Ema:2016ops,Calibbi:2016hwq}, the origin of neutrino masses~\cite{Chikashige:1980ui,Gelmini:1980re,Georgi:1981pg}. 
Besides a possible resolution of the DM problem~\cite{Preskill:1982cy,Abbott:1982af,Dine:1982ah}, the possible connection to other open questions in particle physics makes this class of models very appealing. Furthermore, their rich phenomenology provides a unique way to test new physics at scales inaccessible to any foreseeable collider experiment. 

In this paper, we consider one of the best motivated ALP DM candidates: the majoron, that is, the pNBG associated with the spontaneous breaking of the lepton number~\cite{Chikashige:1980ui}.
This enables us to draw an appealing connection between the origin of neutrino masses, the DM puzzle, past and future searches for DM signatures in photon, electron, and neutrino fluxes from the cosmos, as well as searches for DM-electron scattering at direct-detection experiments such as XENONnT~\cite{XENON:2022ltv}, PandaX~\cite{PandaX:2022ood}, LZ~\cite{Mount:2017qzi}.

After the discovery of neutrino oscillations~\cite{Super-Kamiokande:1998kpq,SNO:2002tuh},
two decades of enormous experimental developments made it possible to measure neutrino masses and mixing with high precision,
see Refs.~\cite{deSalas:2020pgw,Esteban:2020cvm,Capozzi:2021fjo} for the current status.
However, it remains undetermined whether the neutrino mass terms are of the Dirac type, such as all the other SM fermions, or Majorana~\cite{Majorana:1937vz} that would imply that neutrinos and their own antiparticles coincide.
This latter possibility\,\,---\,\,whose simplest realisation is provided by the famous dimension-five Weinberg operator~\cite{Weinberg:1979sa}\,\,---\,\,necessarily requires violation of the lepton number. While this may occur through an explicit breaking, such as the one due to Majorana mass terms of right-handed neutrinos within the simplest realisation of the type-I seesaw mechanism~\cite{Minkowski:1977sc,Gell-Mann:1979vob,Yanagida:1979as,Mohapatra:1979ia}, a spontaneous breaking is also a logical possibility.
In general, the existence of a light majoron would be an unavoidable consequence of any mechanism of generation of neutrino masses involving a spontaneous breaking of the lepton number (as long as the latter is a global symmetry).
 
In short, within a fairly wide class of models, it is natural to add the majoron field to the particle content of the SM alongside the other fields involved in the generation of neutrino masses, thus providing a compelling candidate for ALP DM and a realistic target for searches, among others, of neutrino lines at present and future neutrino telescopes~\cite{Garcia-Cely:2017oco}.

It is also interesting to notice that, in cosmology, majorons have recently gathered attention in the context of possible solutions of the Hubble tension: it has been shown that an additional radiation component due to light majorons as well as neutrino interactions mediated by majorons can relax the tension~\cite{Escudero:2019gvw,Escudero:2021rfi}\,\,---\,\,cf.~Refs.~\cite{Arias-Aragon:2020qip,Fernandez-Martinez:2021ypo,Araki:2021xdk}, for explicit realisations.

In the following pages, we focus in particular on a majoron model~\cite{Choi:1989hj,Choi:1991aa} based on type-II seesaw~\cite{Magg:1980ut,Schechter:1980gr,Cheng:1980qt,Lazarides:1980nt,Mohapatra:1980yp,Wetterich:1981bx}, which consists in adding only two fields to the SM field content: a scalar $SU(2)_L$ triplet and a scalar singlet responsible for the spontaneous breaking of the lepton number. Our aim is to study the model's DM phenomenology that has received comparatively little attention in the literature.\footnote{For early studies on astrophysical signatures of majoron DM within type-I plus type-II seesaw see Refs.~\cite{Bazzocchi:2008fh,Lattanzi:2013uza}. 
Freeze-in production of majoron DM within type-I seesaw has been studied in Ref.~\cite{Frigerio:2011in}.
For a discussion of freeze-in and freeze-out production and experimental consequences of majoron DM within type-I (and type-I plus type-II) seesaw, cf.~Ref.~\cite{Queiroz:2014yna}. See also Ref.~\cite{Chao:2022blc} for a recent study proposing, within the same model we are considering here, a  mechanism of majoron mass generation through explicit breaking plus scalar mixing and the consequent implications for DM production. }
In \sref{sec:typeII}, we review the model and discuss how it is constrained by laboratory experiments, including searches for the triplet at the LHC.
In \sref{sec:DMprod}, we discuss two possible majoron DM production mechanisms, freeze-in and misalignment, both during a standard radiation dominated epoch and an early matter-dominated era. In \sref{sec:constr}, we study the constraints on the parameter space relevant for majoron DM from cosmological surveys, direct detection experiments, and present and future searches for decaying DM at neutrino telescopes and other cosmic-ray probes. We summarise and draw our conclusions in \sref{sec:concl}, while some technical results and useful formulae are collected in the Appendices.

\section{Type-II majoron model}
\label{sec:typeII}

One minimal extension to the SM, which can accommodate Majorana masses of neutrinos, consists in 
introducing an $SU(2)_{L}$ triplet scalar field
with lepton number $L=-2$, which couples with two lepton doublets.
Spontaneous breaking of the lepton number via the vacuum expectation value (vev) of the triplet scalar field induces neutrino Majorana mass terms and, at the same time, provides a Nambu-Goldstone boson, the majoron~\cite{Gelmini:1980re,Georgi:1981pg}.
This minimal scenario for Majorana neutrino masses was already ruled out long ago by the non-observation of majorons in the decays of $Z$ bosons at LEP.\footnote{See Ref.~\cite{Barenboim:2020dmg} for a possible way to evade this constraint introducing higher dimensional operators in the scalar potential.}

The famous type-II seesaw mechanism also features the same scalar triplet~\cite{Magg:1980ut,Schechter:1980gr,Cheng:1980qt,Lazarides:1980nt,Mohapatra:1980yp,Wetterich:1981bx},
but the lepton number is violated explicitly by the scalar cubic interaction involving the triplet and two Higgs doublets. 
Therefore, no majoron appear and 
the phenomenological problems of the triplet scalar model related to the presence of the majoron field are circumvented.
However, one can impose that the origin of the mass-dimensionful
coupling of the scalar cubic term in the type-II seesaw model is 
the vev of an extra scalar field that spontaneously breaks the lepton number.
The new scalar field should be a singlet 
under the SM gauge symmetries but it should be charged under the lepton number $U(1)_L$. 
After the spontaneous breaking of the lepton number, 
a majoron state appears as in the triplet scalar model.
However, this majoron field consists dominantly of the imaginary component of the singlet field and does not suffer the phenomenological problems that affect the original triplet majoron model, being in particular consistent with the measured invisible width of the $Z$ boson.
This possibility, which we call the type-II majoron model, 
was first proposed in Refs.~\cite{Choi:1989hj,Choi:1991aa},
and various aspects of its phenomenology have been discussed in, \emph{e.g.},~Refs.~\cite{Joshipura:1992hp,Diaz:1998zg,Bonilla:2015jdf,Ma:2017xxj}. The model is described in detail in Appendix~\ref{App:typeII}. In this section, we discuss the most relevant aspects for our analysis.

The part of the Lagrangian of the type-II majoron model, 
which is responsible for neutrino masses, reads
\begin{align}
\mathscr{L}_{\text{typeII}}
\,=\,&
(Y_{\Delta})_{\alpha \beta} \,
\overline{L^{c}}_{\alpha}
\Delta
L_{\beta}
+
\kappa \,
\sigma \Phi^{\sf T} \Delta \Phi
+
\text{H.c.}\,,
\label{eq:L_typeII-mNu}
 \end{align}
where $L_{\alpha,\beta}$ are lepton doublets whose flavour is denoted by the indices $\alpha$ and $\beta$,
and $\sigma$, $\Phi$, and $\Delta$ are 
 scalar fields in the $SU(2)_{L}$ singlet, doublet, and triplet representation, respectively.\footnote{%
Here we follow the notation adopted in Ref.~\cite{Bonilla:2015jdf},
where the hypercharge assignment of $\Phi$ is $-1/2$.
The triplet field $\Delta$ contains an anti-symmetric tensor for the index of the $SU(2)_{L}$ fundamental representation, which is necessary to form an $SU(2)_{L}$ singlet together with two $\Phi$s.
See Appendix~\ref{App:typeII} for details.}
The superscript $c$ on the field $L$ denotes charge conjugation.
Notice that both the couplings $Y_{\Delta}$ and $\kappa$ are mass-dimensionless.

The $SU(2)_{L}$ triplet scalar field $\Delta$ carries 
lepton number $L=-2$,
and the singlet field $\sigma$ has $L=+2$, 
while the doublet $\Phi$, which is the SM Higgs scalar field, 
carries no lepton number.
As a consequence, the above Lagrangian is invariant under the $U(1)_L$ lepton number transformation, but that symmetry is spontaneously broken by the vev $\langle \sigma \rangle = v_{1} /\sqrt{2}$ of the singlet field.
This spontaneous breaking of the lepton number brings about a majoron field $J$ stemming from the imaginary part of $\sigma$.

After the electroweak symmetry is broken by the vev of the doublet $\langle \Phi^{0} \rangle = v_{2}/\sqrt{2}$,
the neutral component of the triplet field acquires
a vev $\langle \Delta^{0} \rangle = v_{3}/\sqrt{2}$ too, 
which is mediated by the triplet scalar field itself:
\begin{align}
v_{3} = \frac{1}{2} \kappa \frac{v_{1} v_{2}^{2}}{M_{\Delta}^{2}}\,,
\label{eq:v3-MDelta}
\end{align}
where $M_{\Delta}$ is the mass of the triplet.\footnote{%
Here $M_{\Delta}$ is actually an ``average'' value of the masses of the Higgs bosons from the triplet, and strictly speaking, the triplet vev is mediated by the neutral CP-even Higgs boson $H_{3}$.
The masses of the physical scalar fields,
which are given in Appendix~\ref{App:typeII},
depend on the full description of the scalar potential.
However, in the parameter region that we are interested in, 
the triplet Higgs bosons, $H_{3}$, $A$, $H^{\pm}$,
and $H^{\pm \pm}$, all have similar masses of the order
$M_{\Delta}$, see below.
}
The combination of vevs relevant to the electroweak symmetry breaking is identified as  
\begin{align}
v_{\text{EW}} \equiv
\sqrt{v_{2}^{2} + 2 v_{3}^{2}} ~\simeq ~246\,\text{GeV} \,.
\label{eq:vEW}
\end{align}
The vev of the triplet contributes to 
the masses of the weak gauge bosons and modifies 
the relation between them.
The deviation from the relation predicted by the SM is measured 
in terms of the $\rho$ parameter, which is given by 
\begin{align}
\rho \equiv \frac{M_{W}^{2}}{M_{Z}^{2} \cos^{2} \theta_{W}}
=
\frac{v_{2}^{2} + 2 v_{3}^{2}}{v_{2}^{2} + 4 v_{3}^{2}}=
\frac{v_\textrm{EW}^2}{v_\textrm{EW}^2 + 2 v_3^2}\,.
\label{eq:rho}
\end{align}
The most conservative global fit of the electroweak precision observables (EWPO) reported in Ref.~\cite{ParticleDataGroup:2022pth} gives
$\rho = 1.0002 \pm 0.0009$, 
which sets an upper bound to the triplet vev, 
$v_{3} < 7$~GeV at 95\% confidence level.\footnote{The triplet scalar contributions to the mass of the $W$ boson were revisited in
Refs.~\cite{Kanemura:2022ahw,Heeck:2022fvl,Cheng:2022hbo} 
in light of the recent measurement based on data collected at CDF II~\cite{CDF:2022hxs}. 
Even though large values of $v_{3}$ (as required, {\it e.g.}, by majoron DM production through the freeze-in mechanism, see \sref{sec:DMprod}) decrease $M_{W}$ at the tree level, it is still possible to reproduce the CDF result via loop contributions involving the states in $\Delta$.}

The Yukawa interaction in Eq.~\eqref{eq:L_typeII-mNu}
in combination with the vev of the triplet
provides neutrino with Majorana mass terms:
\begin{align}
(m_{\nu})_{\alpha \beta} 
= - \sqrt{2}\, (Y_{\Delta})_{\alpha \beta}\, v_{3}
=
-\frac{1}{\sqrt{2}} \, (Y_{\Delta})_{\alpha \beta}\, \kappa v_{1} \frac{v_{2}^{2}}{M_{\Delta}^{2}}\,.
\label{eq:mNu-typeII}
\end{align}
Notice that the standard neutrino mass formula of the type-II seesaw 
mechanism is recovered considering that the coupling of the trilinear scalar interaction $\Phi^{\sf T} \Delta \Phi$ is given by \mbox{$\mu = \kappa\, v_{1}/\sqrt{2}$}.


The scalar potential and the masses of the physical states are discussed in \aref{App:typeII}. 
This sector of the model is subject to bounds from vacuum stability and perturbative 
unitarity~\cite{Arhrib:2011uy,Aoki:2012jj,Kannike:2012pe,Bonilla:2015eha,Du:2018eaw}.
In order to have a viable DM candidate,
the vevs of the scalar fields are required to exhibit the hierarchical pattern 
\begin{equation}
\label{eq:vev-hie}
v_{3} \ll v_{2} \ll v_{1}\,.    
\end{equation}
In fact, at least one vev needs to be ultralarge in order to suppress the majoron couplings and make it long-lived enough to be DM (see \sref{sec:constr}) and such a vev can only be $v_1$, because $v_3$ is constrained by EWPO, as discussed above, and $v_{2}$ must provide the main source of the observed electroweak-symmetry breaking.
This implies that the mixing among the CP-odd scalars is very small\,\,---\,\,as one can see by inspecting the mixing matrix $O_I$ in \eref{eq:OI-Opm} for the regime of \eref{eq:vev-hie}. Consistently with this observation, and for the sake of the perturbativity constraints mentioned above, we assume all couplings of the scalar potential to be $\ll 1$ and small mixing to occur in the real scalar sector too. We stress that our conclusions do not depend on the details of the scalar potential, as long as the couplings are not too large. 
The small mixing regime also prevents the majoron (and the real component of the singlet) to thermalise with the hot bath in the early universe.
This and other cosmological consequences of the $\kappa$ term will be discussed in~\sref{sec:DMprod}.
In addition, as we will see below, $\kappa$ is bounded from below by searches for extra Higgs bosons at the LHC.

\subsection{Spectrum of the model and LHC bounds} 
Let us now turn to the spectrum of the model. 
The scalar sector comprises the following physical states:
three CP-even Higgs bosons ($H_{1},\,H_{2},\,H_{3}$),
a CP-odd Higgs boson $A$,
two singly-charged Higgs bosons $H^{\pm}$,
two doubly-charged Higgs bosons $H^{\pm \pm}$,
and the majoron $J$, which\,\,---\,\,together with the three Nambu-Goldstone bosons $(G^0,\,G^\pm)$ that become the longitudinal components of the electroweak bosons\,\,---\,\,add up to the 12 degrees of freedom present in the three complex scalar fields of the model.

Due to the small mixing regime that we are working in,
the (mainly) singlet Higgs boson $H_{1}$ decouples from the rest of the spectrum and is super heavy, $M_{H_{1}} \sim v_{1} \gg v_\textrm{EW}$.
$H_{2}$ consists dominantly of the neutral component of the doublet, and we identify it with the SM Higgs boson, that is, we set $M_{H_{2}}= M_{H_\text{SM}}\simeq 125$~GeV.
The other Higgs bosons mainly originate from the triplet, 
and their masses mainly stem from a common origin, which is 
the $\kappa$ term in Eq.~\eqref{eq:L_typeII-mNu}.
Consequently, they have masses of similar size that, expressed in terms of $\kappa$ 
and the vevs, read
\begin{align}
M_{H_{3}}^{2} 
\simeq 
M_{H_{A}}^{2}
\simeq
M_{H^{\pm}}^{2} 
\simeq
M_{H^{\pm \pm}}^{2}
\simeq
M_{\Delta}^{2} 
\equiv 
\frac{1}{2}
\kappa
\frac{v_{1} v_{2}^{2}}{v_{3}}\,.
\label{eq:MDelta}
\end{align}
For the full description of the scalar sector,
we again refer to Appendix~\ref{App:typeII}.

The collider phenomenology of the triplet Higgs model
has been extensively discussed in, \emph{e.g.}, Refs.~\cite{Chun:2003ej,Akeroyd:2005gt,FileviezPerez:2008jbu,Akeroyd:2011zza,Melfo:2011nx,Aoki:2011pz,Aoki:2012jj,Kanemura:2013vxa,Kanemura:2014goa,Ghosh:2017pxl,Du:2018eaw,Fuks:2019clu,Ashanujjaman:2021txz,Abada:2007ux}.
Since fields charged under the electromagnetic $U(1)$ can be produced through the Drell-Yan process, 
the LHC can set a robust bound to the masses of 
the charged Higgs bosons regardless of the details of the model.
In particular, searches for the production of doubly-charged states $pp\to H^{++} H^{--}$
followed by decays into $W$ bosons, $H^{\pm\pm} \to W^\pm W^\pm$, are the most relevant for scenarios with $v_{3}  > {O}(10^{-4})$~GeV~\cite{Ashanujjaman:2021txz}.
The null result of such searches at the LHC~\cite{ATLAS:2021jol}
provide a lower limit on the mass of the triplet,
$M_{\Delta} \gtrsim 400$~GeV. Following from \eref{eq:MDelta}, this limit can be interpreted in our case as a lower bound to $\kappa$: 
\begin{align}
 \kappa 
 \gtrsim 
 2.7 \cdot 10^{-7}
 \left[
 \frac{M_{\Delta}}{400\,\text{GeV}}
 \right]^{2}
 \left[
 \frac{v_{3}}{1\,\text{GeV}}
 \right]
 \left[
 \frac{2 \cdot 10^{7}\,\text{GeV}}{v_{1}}
 \right].
\label{eq:kappa-MD-v1-v3}
\end{align}

Finally, we need to discuss the mass $M_J$ of the majoron field.
Indeed, the majoron can not be a genuine Nambu-Goldstone boson\,\,---\,\,that is, exactly massless\,\,---\,\,in order for it to be a candidate for DM. In other words, a majoron mass term must arise out of a (small) explicit breaking of the lepton number. Such breaking can originate from higher-dimensional Planck-suppressed operators, which is consistent with the commonly accepted notion that no global symmetry is preserved within quantum gravity\,\,---\,\,see, {\it e.g.}, Ref.~\cite{Reece:2023czb} for a recent review. 
A straightforward example is provided by operators only involving the scalar singlet~\cite{Akhmedov:1992hi,Rothstein:1992rh}:
\begin{align}
\mathscr{L}_\textsc{lnv}
\supset  \frac{\lambda_{m,n} }{M_\textrm{Pl}^{d-4}} \sigma^m \sigma^{*\,n} +
\text{H.c.}\,,
\label{eq:Planck-ops}
\end{align}
where $M_{\text{Pl}}=1.22\cdot 10^{19}$ GeV is the Planck mass,
$d=m+n \ge 5$, $\lambda_{m,n}$ are dimensionless coefficients, and the lepton number is explicitly broken if $m\neq n$.
As one can see, the above operators would induce contributions to the majoron mass $M_J$ as
\begin{align}
M_J^2
\sim  \lambda_{m,n} \frac{v_1^{d-2}}{M_\textrm{Pl}^{d-4}}\,.
\end{align}
As we will show in the following, majoron DM requires $v_1 \gtrsim 10^{7}$~GeV. In such a regime, $d=5$ operators induce 
\begin{align} 
M_J \simeq 10~\text{GeV} \left(\lambda_{m,n}\right)^{1/2} \left(\frac{v_1}{10^7~\text{GeV}}\right)^{3/2}\,,
\end{align}
thus the mass range relevant for DM detection ($10~\text{keV}\lesssim M_J \lesssim 100~\text{MeV}$, see \sref{sec:constr}) require small couplings $\lambda_{m,n} \lesssim 10^{-12} - 10^{-4}$ for $v_1\gtrsim 10^7$~GeV. If the dominant contributions to $M_J$ arise at dimension 6, we have instead
\begin{align} 
M_J \simeq 10~\text{keV} \left(\lambda_{m,n}\right)^{1/2} \left(\frac{v_1}{10^7~\text{GeV}}\right)^{2}\,,
\end{align}
hence the whole of the mass range we are interested in is accessible for $d=6$ operators with coefficients $\lambda_{m,n} = \mathcal{O}(1)$.

Notice that, even if present, operators such as $\sigma^m \sigma^{*\,n} \left(\Phi^\dag \Phi\right)^p$ and $\sigma^m \sigma^{*\,n} \left(\Delta^\dag \Delta\right)^p$ (of dimension $d =m+n+2\,p$, with $m+n \ge 3$ and $p\ge 1$) would give subdominant contributions to $M_J$ compared to the operators in Eq.~(\ref{eq:Planck-ops}) of the same dimensionality, as a consequence of the vev hierarchy of \eref{eq:vev-hie}. 
As we will see in the next subsection, the couplings of the majoron relevant to DM phenomenology (that is, those to photons and SM fermions) arise from mixing in the scalar sector. As a consequence, the explicit breaking of the lepton number provided by Eq.~(\ref{eq:Planck-ops})\,\,---\,\,which does not substantially contribute to the mixing among singlet, doublet and triplet\,\,---\,\,does not affect the majoron couplings and thus DM production and decays.\footnote{At first order in perturbation theory, the majoron mass term introduces the following correction to the mass eigenstates of~\eref{eq:masseig}: 
\begin{align*}
J^{(1)} &= J^{(0)} - \frac{(\delta M^2)_{31}}{M_A^2} A^{(0)} + \frac{(\delta M^2)_{23} (\delta M^2)_{31}}{M_J^2 M_A^2} G^{(0)} \simeq J^{(0)} - \frac{M_J^2}{M_A^2}  \frac{v_3}{v_1}  A^{(0)} \,, \nonumber\\
G^{(1)} &= G^{(0)} - \frac{(\delta M^2)_{32}}{M_A^2} A^{(0)} + \frac{(\delta M^2)_{13} (\delta M^2)_{32}}{M_J^2 M_A^2} J^{(0)} = G^{(0)} \,,\nonumber\\
A^{(1)} &= A^{(0)} + \frac{(\delta M^2)_{13}}{M_A^2} J^{(0)} + \frac{(\delta M^2)_{23}}{M_A^2} G^{(0)}  \simeq A^{(0)} + \frac{M_J^2}{M_A^2}  \frac{v_3}{v_1}  J^{(0)}\,,
\end{align*}
where $(\delta M^2)_{ij} \equiv (O_I)_{i1} M_J^2 (O_I^{\sf T})_{1j}$ and we considered the mixing matrix $O_I$ of \eref{eq:OI-Opm} in the regime $v_3 \ll v_2 \ll v_1$.
As we can see, for $M_J \lesssim 100$~MeV, $M_A \simeq M_\Delta > 100$~GeV, the operators responsible for $M_J^2$ induce a relative correction $\sim M_J^2 / M_A^2 \lesssim 10^{-6}$, thus completely negligible, to the singlet-triplet mixing (which is $\sim v_3/v_1$ at leading order) and do not affect the mixing with the doublet.
}
Hence $M_J$ can be safely treated as a free parameter with no consequence on the majoron interactions, and so will we do throughout the paper.\footnote{This obviously requires exploiting the freedom of adjusting the above-defined operator coefficients $\lambda_{m,n}$.}

\subsection{Majoron couplings and decays}
\label{sec:Jcoupl}

Being dominantly part of the singlet $\sigma$, the type-II majoron inherits couplings to SM fields through mixing in the scalar sector, as shown in \aref{App:typeII}. These can be written as
\begin{align}
\mathscr{L}_{J}
=
{\rm i}
g_{Jff}^{P}\,
J\,
\overline{f} \gamma^{5} f
-
\frac{1}{4}
g_{J \gamma \gamma}
J\, F_{\mu \nu} \widetilde{F}^{\mu \nu}\,,
\end{align}
where $f$ denotes the SM fermion species: neutrinos, charged leptons, up and down quarks, \mbox{$f = \nu,\,\ell,\,u,\,d$}.
The couplings to the neutrino mass eigenstates ($\nu_i$, $i=1,2,3$) stem from the interaction $Y_\Delta$ of the triplet with lepton doublets in \eref{eq:L_typeII-mNu} and, for the hierarchical regime $v_3\ll v_2 \ll v_1$ that we are considering, read 
\begin{gather}
g_{J \nu_{i} \nu_{i}}^{P}
\simeq
- \frac{m_{\nu_{i}}}{2 v_1}\,.
\end{gather}
The couplings to the other fermions are
\begin{gather}
g_{J \ell_{\alpha} \ell_{\alpha}}^{P}
\simeq
-
 m_{\ell_{\alpha}} \, \frac{2 v_{3}^{2}}{v_1 v_2^2}\,,
\quad
g_{J u_{\alpha} u_{\alpha}}^{P}
\simeq
 m_{u_{\alpha}}\, \frac{2 v_{3}^{2}}{v_1 v_2^2}\,,
\quad
g_{J d_{\alpha} d_{\alpha}}^{P}
\simeq
-
 m_{d_{\alpha}}\, \frac{2 v_{3}^{2}}{v_1 v_2^2}\,,
\label{eq:Jcoupl}
\end{gather}
 where $\alpha =1,2,3$ denotes the flavour of the fermion and, again, we restricted to the hierarchical regime shown in \eref{eq:vev-hie}\,\,---\,\,for the full expressions beyond this limit, see \aref{App:typeII}. Notice that the couplings to quarks and charged leptons are $\propto v_3^2/v_2^2$. Furthermore, they are flavour-conserving, since they are all inherited from the fermion interactions with the Higgs doublet\,\,---\,\,in fact, $\Delta$ does not couple to quarks and its neutral component does not couple to charged leptons either\,\,---\,\,hence rotations of the fields to the mass basis diagonalise them too.

Finally, the effective coupling with photons arises from majoron-pion mixing as well as loops of charged fermions:
\begin{gather}
g_{J\gamma \gamma}
\simeq
\frac{2\alpha}{\pi}
  \frac{v_{3}^{2}}{v_1 v_2^2}
\left[ 
\frac{M_{J}^{2}}{M_{J}^{2} - m_{\pi^{0}}^{2}}
- 
\sum_f Q_f^2 N_c^f
B_{1}(\tau_{f})
\right],
\label{eq:g_Jphph-typeII}
\end{gather}
where $Q_f$ and $N^f_c$ are, respectively, electric charge and colour multiplicity of the fermion $f$, and \mbox{$\tau_{f} \equiv 4 m_{f}^{2}/ M_{J}^{2}$}.
The expression of the loop function $B_{1}(\tau_f)$ can be found in \aref{App:typeII}. This function is such that $B_{1}(\tau_f)\to 0$ for $\tau_f \to \infty$, that is, the majoron decouples from photons for $M_J \ll m_e$. This is because the lepton number is anomaly free hence there is no anomalous majoron coupling with photons.\footnote{ALP couplings to the weak gauge bosons are similarly induced, cf.~\emph{e.g.}~Ref.~\cite{Bauer:2020jbp}. However, in the case of the type-II majoron within the regime in \eref{eq:vev-hie}, they are too small to give rise to observable effects, such as the collider observables discussed in Ref.~\cite{Brivio:2017ije} for general ALPs.}

In the mass range relevant to DM direct detection and astrophysical probes that will focus on in \sref{sec:constr}, that is, \mbox{$M_J\approx$ 1\,keV\,--\,100\,MeV},
the majoron can only decay into photons, neutrinos, and possibly electrons.
The corresponding decay rates read (see, {\it e.g.},~\mbox{Refs.~\cite{Heeck:2019guh,Calibbi:2020jvd}})
\begin{align}
\label{eq:Jgammas}
\Gamma(J \rightarrow \gamma \gamma)
~=~&
\frac{M_{J}^{3}}{64 \pi} 
\left|
g_{J \gamma \gamma} 
\right|^{2},\\
\label{eq:Jnunu}
\Gamma(J \rightarrow \nu_{i} \nu_{i})
~=~&
\frac{M_{J} }{4 \pi} \left|
g_{J \nu_{i} \nu_{i}}^{P}
\right|^{2}
~\simeq~
\frac{M_{J} }{16 \pi}  \left(\frac{m_{\nu_{i}}}{v_1}\right)^2,  \\
\label{eq:Jee}
\Gamma(J \rightarrow e^+ e^-)
~=~&
\frac{M_J}{8 \pi} 
\left|
g^P_{J e e} 
\right|^{2} \,\sqrt{1-\frac{4m_e^2}{M_J^2}}
~\simeq~ \frac{M_J}{2 \pi} 
\left(\frac{m_e}{v_1 }\right)^2  \left(\frac{v_3}{v_2}\right)^4
\,\sqrt{1-\frac{4m_e^2}{M_J^2}}\,.
\end{align}
Cosmological and astrophysical consequences of these decays are discussed in \sref{sec:constr}.


\subsection{Other laboratory constraints}

The type-II seesaw triplet can mediate charged-lepton-flavour-violating (cLFV) processes, such as $\mu \to ee\bar e$, cf.~for instance Ref.~\cite{Abada:2007ux}. See also Ref.~\cite{Calibbi:2022wko} for a discussion of distinctive cLFV signals of type-II seesaw within the context of grand unified theories.
The current limit $\text{BR}(\mu \to ee\bar e) < 10^{-12}$~\cite{Bellgardt:1987du} translates into a bound $M_\Delta /\sqrt{|(Y_\Delta)_{21}||(Y_\Delta)_{11}|} > 200$~TeV, hence
cLFV muon decays typically require $(Y_{\Delta})_{\alpha\beta} \lesssim 0.005$ for $M_{\Delta} \sim 1$~TeV.
As we will discuss in the next section,  freeze-in production of majoron DM is only possible if indeed $M_\Delta \lesssim 1$~TeV.
However, we have seen that neutrino masses require $Y_{\Delta} \sim m_\nu/v_3$, hence we can not expect a signal at the upcoming cLFV search campaign (see, \emph{e.g.},~\cite{Calibbi:2017uvl}) unless $v_3 \lesssim 10^{-7}$~GeV, a regime where the majoron DM phenomenology is solely controlled by its decays into neutrinos, freeze-in DM production is not effective and the observed relic density can only be accounted for by the misalignement mechanism, cf.~Sections~\ref{sec:DMprod} and \ref{sec:constr}.
In addition, cLFV decays into majorons, such as $\mu\to e J$, which can be powerful probes of the leptonic couplings of other kinds of majorons as well as generic ALPs\,\,---\,\,cf.~Refs.~\cite{Calibbi:2020jvd,Jho:2022snj} for an overview\,\,---\,\,can not occur for the type-II majoron, because its couplings to charged leptons are \emph{flavour conserving}, as discussed above\,\,---\,\,see \eref{eq:Jcoupl}.

Neutrinoless double beta decays can also be mediated by the doubly-charged Higgs boson, which is a component of the triplet scalar~\cite{Mohapatra:1981pm}.
The size of the contribution,
given in terms of the vevs and the triplet 
mass $M_{\Delta}$, is approximately 
${m_{e} v_{3}^{3}}/({v_{1} v_{2}^{2} M_{\Delta}^{2}})$.
This has to be compared with the standard 
Majorana neutrino mass contribution 
$(m_{\nu})_{ee}/p_{\nu}^{2}$,
where $p_{\nu}$ is the momentum carried by 
a neutrino in an atom, which is
typically of the order of $0.1$~GeV.
Considering the values of the parameters
required by majoron DM, in particular $v_1 \gtrsim 10^7$~GeV (see \sref{sec:constr}),
one can find that the triplet contribution is 
smaller than the standard one (setting $(m_{\nu})_{ee}$ to a value compatible with the current limit from Ref.~\cite{KamLAND-Zen:2022tow}) by several orders of magnitude. 
In addition, a recent result of the search 
for neutrinoless double beta decays 
with majoron emission provides a direct bound to
the majoron-neutrino coupling at the level of $|g_{J\nu_{e} \nu_{e}}| \lesssim 10^{-5}$~\cite{Kharusi:2021jez},
which is also significantly larger than the size 
of the coupling in the parameters' range relevant for DM majorons,
\mbox{$|g^P_{J\nu_i \nu_i}| \simeq m_{\nu_i} /(2 v_1) \lesssim 10^{-18}$}.

Majoron-quark interactions can be constrained by searches for $K^{+} \rightarrow \pi^{+} + $\,invisible 
at kaon experiments such as NA62~\cite{NA62:2021zjw}.
In our case, 
the dominant contribution to processes of the kind $K\to \pi J$ comes from a $J$-penguin diagram with a top and a $W$ in the loop~\cite{Bezrukov:2009yw}. 
See also Ref.~\cite{Guerrera:2021yss}, where this process was recently revisited.
The authors of Ref.~\cite{Bezrukov:2009yw} discuss the decay occurring through the mixing of the Higgs with a light scalar $\chi$ and give 
a bound to the mixing parameter $\theta$ defined as 
$(\theta m_{f}/v) \chi \bar{f} f$.
In the type-II majoron model, 
the mixing parameter $\theta$ then corresponds to 
$\sqrt{2} v_{3}^{2}/(v_{1} v_{2})$, see \eref{eq:Jcoupl}.
According to Ref.~\cite{Guerrera:2021yss}, the NA62 limit translates into the bound $\theta \lesssim 10^{-4}$,
while for values of the vevs corresponding to a majoron DM we find  $\theta \lesssim 10^{-8}$.
Similarly, we find that the parameter region where the majoron is a good DM candidate is far beyond the sensitivity reach of ALP searches at B-factories~\cite{Merlo:2019anv} and in $\pi$ decays~\cite{Altmannshofer:2019yji}.

\section{Majoron production in the early universe}
\label{sec:DMprod}

Throughout this paper, we require that the parameters of the model, besides  fulfilling the constraints discussed in the previous section, be such that the majoron is a good DM candidate.
To be more specific, the majoron must be produced in the early universe
as cold DM, and its relic abundance today must reproduce the value inferred from cosmological observations, $\Omega_{\rm DM} h^{2} = 0.12$~\cite{Planck:2018vyg}. 
In this section, we review two possible production mechanisms for majoron DM and provide the formulae to estimate its relic density.

\subsection{Misalignment mechanism}
\label{sec:misalignment}

The misalignment mechanism, first introduced for the QCD axion~\cite{Preskill:1982cy,Abbott:1982af,Dine:1982ah}, 
is one of the most popular DM production mechanisms within ALP scenarios in general~\cite{Arias:2012az}. If the global symmetry (in our case, the lepton number) is broken before inflation, the ALP field attains a uniform value across all observable universe.
If such initial value is misaligned from the minimum of the potential, the field starts falling to the minimum at the temperature $T_{\text{osc}}$ where the Hubble parameter $H(T)$ becomes comparable to the mass of the ALP, and the energy dissipated by the oscillation of the classical field about the minimum is converted into energy carried by the ALP, such that the resulting relic ALPs act as cold DM.

\begin{figure}[t!]
\centering
\includegraphics[width=0.55\textwidth]{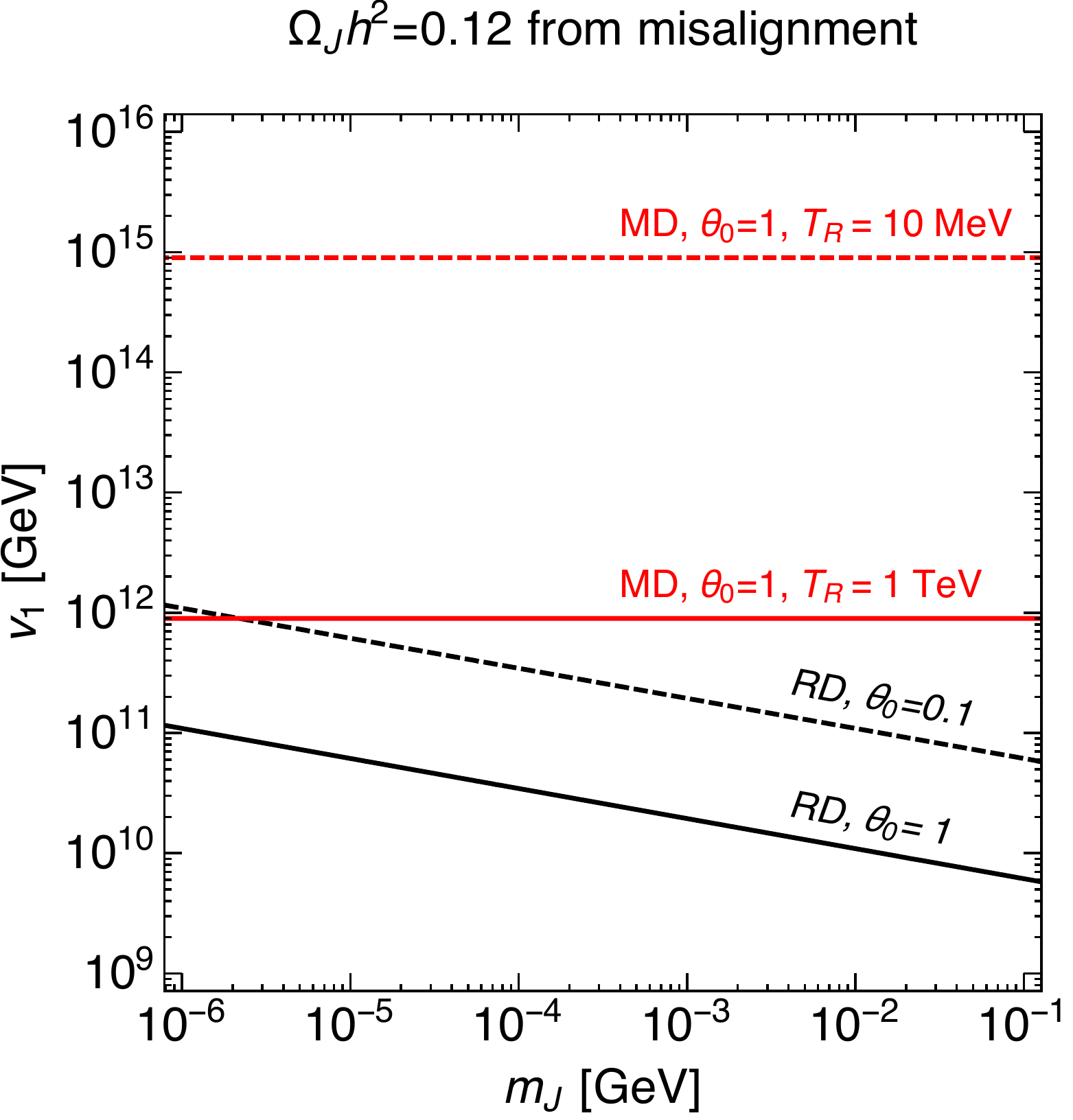}
\caption{Contours on the $M_J-v_1$ plane corresponding to $\Omega_J h^2 = 0.12$ from the misalignment mechanism for majoron oscillations starting either during a standard radiation-dominated era (RD, black lines) or an early matter-dominated era (MD, red lines), and for different values of the initial misalignment angle $\theta_0$ and the reheating temperature $T_R$.
}
\label{Fig:misalign}
\end{figure}

Assuming a standard evolution of the universe (with radiation dominating the energy budget of the universe when the field starts oscillating), one obtains the following estimate of the majoron relic density~\cite{Blinov:2019rhb} (see also Ref.~\cite{Arias:2012az}) 
\begin{align}
\Omega_{J} h^{2}
=
\frac{\rho_{J}(T_{\text{osc}})}
{\rho_{c0}}
\frac{g_{*s}(T_{0})\, T_{0}^{3}}
{g_{*s}(T_{\text{osc}})\, T_{\text{osc}}^{3}} h^{2},
\label{eq:OmegahSq-misalignment}
\end{align}
where 
$h\simeq 0.68$,\footnote{The measurements of the local distance ladder suggest a larger value, cf.~the review in Ref.~\cite{ParticleDataGroup:2022pth}. Here we adopt the value suggested by the Planck satellite~\cite{Planck:2018vyg}.}
$\rho_{c 0}$ is the critical density today,
$g_{*s}(T)$ is the effective number of entropy degrees of freedom at a temperature $T$,
$\rho_{J}(T)$ is the energy density of majoron at $T$,
and $T_{0}$ is the temperature of radiation (that is, relic photons) today.
With the initial misalignment of the  majoron field parameterised as $J_{0} = v_{1} \theta_{0}$, the energy density at $T_{\text{osc}}$ is expected to be
$
\rho_{J}(T_{\text{osc}}) = 
\frac{1}{2} M_{J}^{2} v_{1}^{2} \theta_{0}^{2}
$.
Substituting this into Eq.~\eqref{eq:OmegahSq-misalignment}, we find that the final relic density would be determined by the initial value of the majoron field, of the order of the scale of the lepton number violation as~\cite{Blinov:2019rhb}
\begin{align}
\Omega_{J} h^{2}
\simeq
0.12
\left[
\frac{v_{1} \theta_{0}}
{1.9 \cdot 10^{13}\,\text{GeV}}
\right]^{2}
\left[
\frac{M_{J}}{1\,\text{$\mu$eV}}
\right]^{1/2}
\left[
\frac{90}{g_{*}(T_{\text{osc}})}
\right]^{1/4},
\label{eq:OmegahSq-standard-misalign}
\end{align}
where $g_{*}(T_{\text{osc}})$ is the effective number of relativistic degrees of freedom at $T_\text{osc}$.
In \fref{Fig:misalign}, we show the contours corresponding to $\Omega_J h^2 = 0.12$ as black lines on the $M_J-v_1$ plane, assuming a standard history of the early universe.
As we will see in \sref{sec:constr}, direct DM searches through electron recoil events, such as those performed at the XENONnT experiment~\cite{XENON:2022ltv}, can be sensitive to  majoron DM with $M_{J} \sim \mathcal{O}(1-10)$ keV, $v_{1} \sim \mathcal{O}(10^{7})$ GeV, and $v_{3} \sim \mathcal{O}(1)$ GeV.
For those values of the parameters, the formula in Eq.~\eqref{eq:OmegahSq-standard-misalign} suggests that the standard misalignment mechanism could only account for a tiny fraction of the observed DM relic density.
As a consequence, if a DM signal will be discovered at XENONnT or other direct searches in the future, one would need another production mechanism.
On the other hand, we will also see that searches for DM decaying into gamma rays or neutrinos can be sensitive to majoron DM produced through the misalignment mechanism.

Since the history of the universe before the big bang nucleosynthesis (BBN) is not experimentally probed yet, we are still allowed to assume various non-standard cosmological scenarios to occur between inflation and the BBN.
For example, it is possible that the universe underwent an era in which the energy budget was dominated by non-relativistic matter before the radiation-dominated era, so long as the reheating temperature $T_{R}$ (at which the two eras are switched) is set to be higher than the BBN temperature, which is in the MeV range~\cite{Hannestad:2004px,DeBernardis:2009cvl,Kawasaki:2000en,Ichikawa:2005vw}.
In such a setup, the oscillation of the majoron field would start during the early matter-dominated era.
The formula for the relic density in Eq.~\eqref{eq:OmegahSq-misalignment} has to be modified by replacing $T_{\text{osc}}$ with the reheating temperature $T_{R}$.
The energy density at $T_{R}$ is related to that at $T_{\text{osc}}$ as $\rho_{J}(T_{R}) = \rho_{J} (T_{\text{osc}}) H^{2}(T_{R})/H^{2}(T_{\text{osc}})$, since the majoron is cold DM and $H^{2} \propto a^{-3}$ in the early matter-dominated era.
The initial energy density is as in the standard case, that is,
$\rho_{J}(T_{\text{osc}}) = \frac{1}{2} M_{J}^{2} v_{1}^{2} \theta_{0}^{2} \simeq v_{1}^{2}\theta_{0}^{2} H^{2}(T_{\text{osc}})$.
Substituting these expressions into the modified Eq.~\eqref{eq:OmegahSq-misalignment},
we can find that the final relic density is~\cite{Blinov:2019rhb}
\begin{align}
\Omega_{J} h^{2}
\simeq 0.12
\left[
    \frac{v_{1} \theta_{0}}{9 \cdot 10^{14}\,\text{GeV}}
\right]^{2}
\left[  
    \frac{T_{R}}{10\, \text{MeV}}
\right],
\end{align}
which, as we can see, does not depend on the mass of the majoron nor on $g_{*s}$.
The contours corresponding to $\Omega_J h^2 = 0.12$ for early matter domination are shown as red lines in \fref{Fig:misalign}.
Reproducing the correct relic density with $v_{1} \sim \mathcal{O}(10^{7})$ GeV would require too high a value of $T_{R}$, in fact inconsistently  $T_{R} \gg T_\text{osc}$. Hence the misalignment mechanism in the early matter-dominated era would do not help explain a possible DM signal at the XENONnT experiment in the future\,\,---\,\,however, it further opens up the parameter space that can be tested by indirect DM searches, as we will see in the next section. 
As shown in the following, majoron DM with such low values of $v_{1}$ can be instead efficiently produced via the freeze-in mechanism.

\subsection{Freeze-in mechanism}
\label{sec:FI}

As we have seen in the previous section,
 LHC searches for extra Higgs bosons require
non-zero $\kappa$, 
cf.~Eq.~\eqref{eq:kappa-MD-v1-v3}, 
and therefore, 
there are inevitable scalar-majoron interactions, which 
 lead to majoron production in the early universe, even if the majoron field is decoupled from the thermal bath due to its feeble couplings, through the so-called freeze-in mechanism~\cite{Hall:2009bx}.
The possible production processes are\,\footnote{
Here we assume $M_{H_{3}} > M_{A}$.
If $M_{A} < M_{H_{3}}$, 
$A \rightarrow H_{3} J$ is counted in, 
instead of $H_{3} \rightarrow A J$.
}
\begin{align}
H^{\pm} \rightarrow W^{\pm} J\,,
\quad
H_{2,3} \rightarrow Z J\,,
\quad
A \rightarrow H_{2} J\,,
\quad
H_{3} \rightarrow A J\,,
\quad 
H_{2,3} \rightarrow JJ\,.
\label{eq:FI-decay-process}
\end{align}
The decay rates of these processes and the details of the relic density calculation are given in Appendix~\ref{App:RelicDensity}.
We find that the majoron pair-production processes 
are suppressed in comparison with the single production ones,
and
the rates of the decays of the SM-like Higgs boson~$(H_{2})$ 
are smaller than those of a triplet Higgs boson~$(H_{3}, A, H^{\pm})$ by orders of magnitude.
Furthermore, the process $H_{3} \rightarrow A J$ is suppressed 
by the small mass splitting between $H_{3}$ and $A$. 
Therefore, the dominant production processes are the first three in Eq.~\eqref{eq:FI-decay-process} with 
the triplet scalar in the initial state.

Freeze-in production is at work approximately until the temperature of the universe drops below the mass of the parent particle, which is $M_{\Delta}$ in this scenario.
Note that before the electroweak phase transition, 
the vevs $v_{2}$ and $v_{3}$ are null and 
the majoron interactions are turned off,
\emph{i.e.},
the freeze-in production happens only after the EWPT.
In other words, to make the triplet scalars massive and turn on 
the cubic interactions with a majoron 
at the temperature $T > M_{\Delta}$,
we have to assume that the electroweak phase transition (EWPT) occurs at a temperature higher than 
the triplet scalar mass, \emph{i.e.},~$T_{\textsc{ewpt}} > M_{\Delta} > 400$ GeV.\footnote{%
In our calculations of the majoron relic density, we do not specify the temperature of the EWPT and use the standard freeze-in formulae with an integration of the majoron production from an infinitely high temperature. 
However, the total relic density is dominated by the result of the integration around $T \sim M_{\Delta}$.
Therefore, our results are still valid if one assumes that $T_{\text{EWPT}}$ is higher than 
$M_{\Delta}$ only by an $\mathcal{O}(1)$ factor.}
Naively, the critical temperature of EWPT in the SM is 
expected to
be around the mass of the SM Higgs boson~\cite{DOnofrio:2015gop}, 
and therefore,
freeze-in majoron production seems to require an extension 
of the scalar potential, such that the critical temperature is increased.
In this work, we do not step into the issue of the phase transition. Nevertheless, we notice the the EWPT at a temperature higher than the SM Higgs mass is not forbidden by any phenomenological observation.
For recent studies on high-temperature EWPT (and EW symmetry non-restoration) see, \emph{e.g.},~Refs.~\cite{Meade:2018saz,Baldes:2018nel,Biekotter:2021ysx,Carena:2021onl}.

Employing Eqs.~\eqref{eq:v3-MDelta}, \eqref{eq:main-freezein}, and \eqref{eq:OmegahSq-FI-BBJ},
the relic density from the above decay processes can be estimated as\,\footnote{The scattering processes also contribute to majoron production. However, they are always sub-dominant within the parameter space that we are interested in, see Appendix~\ref{App:RelicDensity}.} 
\begin{align}
\Omega_{J} h^{2}
\simeq
0.12
\left[\frac{110}{g_{*}(M_{\Delta})}\right]^{3/2}
\left[
\frac{M_{J}}{10\,\text{keV}}
\right]
\left[
\frac{M_{\Delta}}{500\,\text{GeV}}
\right]
\left[
\frac{2\cdot 10^{9} \,\text{GeV}}{v_{1}}
\right]^{2}
\left[
\frac{v_{3}}{5\,\text{GeV}}
\right]^{2},
\label{eq:OmegahSq-standard-FI}
\end{align}
such that the majoron is overproduced in the parameter region explored by direct detection experiments\,\,---\,\,$M_{J} \sim \mathcal{O}(1-10)$~keV, 
$v_{1} \sim \mathcal{O}(10^{7})$~GeV,
$v_{3} \sim \mathcal{O}(1)$~GeV\,\,---\,\,see Section~\ref{sec:constr}.

Majoron overproduction is not the only problem.
It is possible that the scalar quartic interactions from the $\kappa$ term keep the majoron in equilibrium with the thermal bath, in contrast with the starting assumption of the freeze-in mechanism.
To avoid thermal equilibrium (and thermal production) of majorons, we have to set the value of $\kappa$\,\,---\,\,or equivalently that of $M_{\Delta}$ once the vevs are fixed, see \eref{eq:MDelta}\,\,---\,\, smaller than a certain value.
This upper bound to $\kappa$ can be estimated through the Gamow's criterion of thermalisation,
\emph{i.e.}, comparing the rate $\Gamma$ of a given process with the rate $H$ of expansion of the universe.
To be specific, we focus on the following 
quartic interaction contained in the $\kappa$ term of Eq.~\eqref{eq:L_typeII-mNu},
\begin{align}
\mathscr{L}_{\text{typeII}}
\supset 
\frac{1}{2} \kappa I_{1} R_{2}^{2} I_{3}\,,
\end{align}
where $I_{1}$ is the imaginary part of the singlet scalar field, $R_{2}$ is the real part of the neutral component of the doublet field, and
$I_{3}$ is the imaginary component of the triplet field.
As discussed in Section~\ref{sec:typeII}, the majoron field consists dominantly of $I_{1}$, and
$R_{2}$ and $I_{3}$ are the main components of 
the SM Higgs boson $H_{2}$ and 
the CP-odd Higgs boson $A$, respectively.  
In short, this interaction induces majoron production through the process 
$H_{2} H_{2} \rightarrow A J$.
The rate $\Gamma(\kappa,T)$ of this scattering process is estimated as  
\begin{align}
\Gamma ({\kappa},T) \equiv 
n_{R_{2}} (T)
\sigma(R_{2} R_{2} \rightarrow I_{3} I_{1}; \kappa,T)\,,
\end{align}
where $n_{R_{2}}$ is the number density of $R_{2}$ 
and $\sigma(R_{2} R_{2} \rightarrow I_{3} I_{1})$
is the cross section of the process,
which is given by
$\sigma = |\kappa|^{2}/(16 \pi E_{\text{CM}}^{2})$
in the massless limit.
The centre-of-mass energy $E_{\text{CM}}$ 
at a given temperature $T$ 
can be estimated as $E_{\text{CM}} = 2 E_{R_{2}} 
\simeq 2 \rho_{R_{2}} / n_{R_{2}} \simeq 5.4\,T$,
where
the number and energy densities, $n_{R_{2}}$ and 
$\rho_{R_{2}}$, of $R_{2}$ are given by the 
Bose-Einstein distribution.
The thermalisation is assessed by comparing this interaction rate $\Gamma$ with the Hubble parameter $H(T)$ at a given $T$.
If $\Gamma(\kappa, T) < H(T)$, the interaction stays out of equilibrium at $T$.
The Hubble parameter in the standard radiation-dominated early universe is given by
\begin{align}
H(T) = \sqrt{\frac{4 \pi^{3}}{45} g_{*}(T)} \frac{T^{2}}{M_{\text{Pl}}},
\end{align}
where $M_{\text{Pl}}=1.22\cdot 10^{19}$ GeV is the Planck mass, and the number of degrees of freedom can be set at $g_{*} \simeq 110$ in the range of temperatures we are interested in.

\begin{figure}[t!]
\centering
\includegraphics[width=0.6\textwidth]{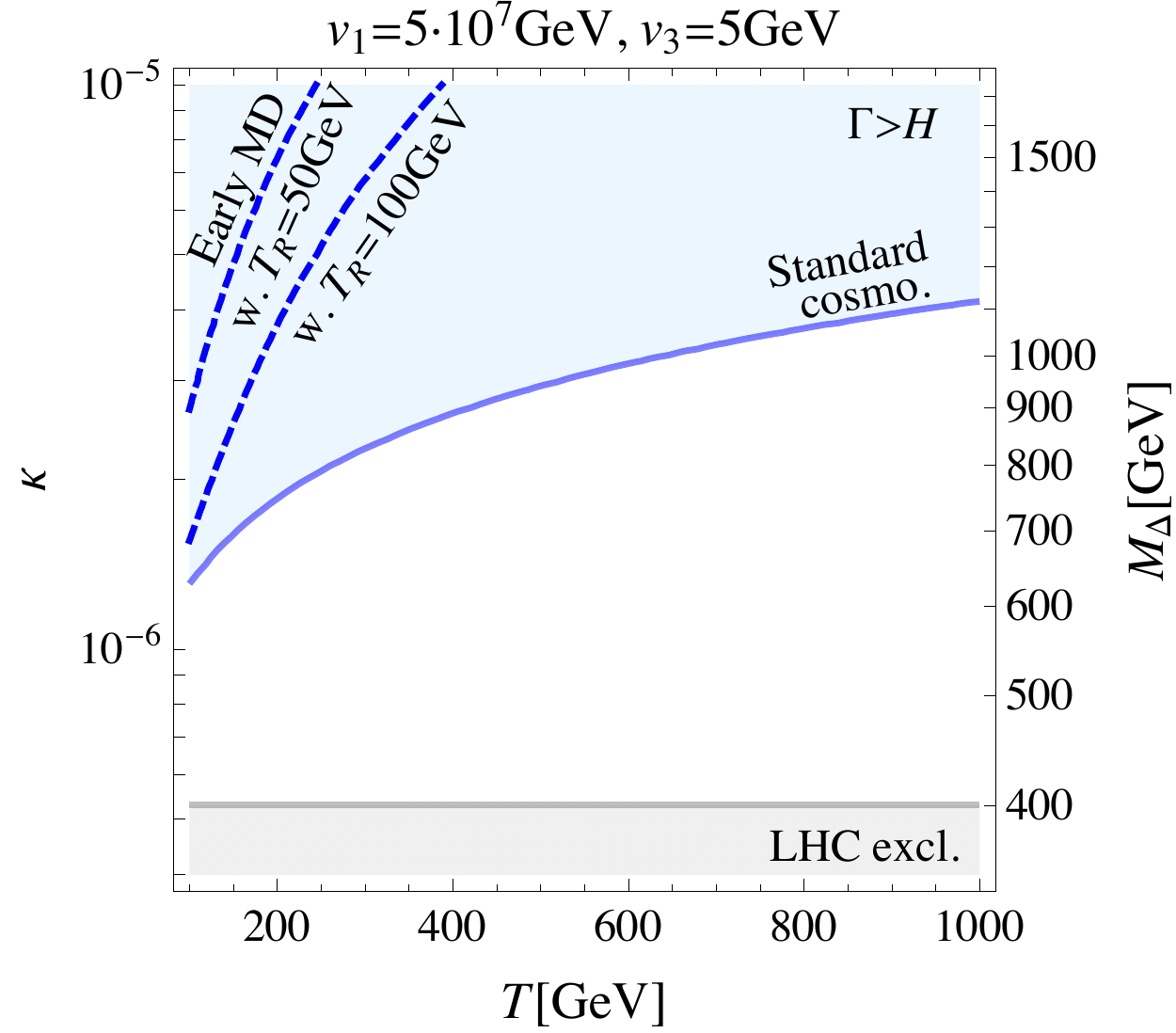}
\caption{Constraints on $\kappa$:
temperature at which a four-scalar interaction in the $\kappa$ term enters thermal equilibrium is indicated with the solid blue curve, \emph{i.e.},~majorons are thermalised through the $\kappa$ interaction in the light blue region.
The extra Higgs searches at LHC set the lower bound to $M_{\Delta}$, which is indicated with a gray line~\cite{Ashanujjaman:2021txz}.
By assuming an early matter-dominated era and setting 
the reheating temperature as 
$T_{R} = \{50,100\}$ GeV,
the solid blue curve of the thermalisation condition
is relaxed as indicated by the dashed curves.
}
\label{Fig:kappa-equilibrium}
\end{figure}

In \fref{Fig:kappa-equilibrium},
the boundary of the condition, $\Gamma = H$, is 
shown in the $T$-$\kappa$ plane with a solid blue curve,
where $v_{1}$ and $v_{3}$ are set at values compatible with the sensitivity of XENONnT, as discussed in Section~\ref{sec:constr}.
Note that since $\Gamma/H \propto 1/T$,
the scattering process comes into equilibrium
only at a low temperature.
This plot should be read as follows;
Fixing a value of $\kappa$, \emph{i.e.},~$M_{\Delta}$,
we determine a horizontal line.
Then, we start from the right end (high $T$) of 
the line and move to the left (low $T$) along 
the line as the universe evolves.
When we hit the blue curve, 
the process $H_{2}H_{2} \rightarrow AJ$ 
comes into equilibrium and majorons are thermally produced 
through it.
To avoid thermal production of majorons through the scattering process, 
the interaction rate $\Gamma$ must be suppressed
before we hit the blue curve.
There are two possibilities to circumvent the thermalisation:
(a)~$n_{R_{2}}$ is suppressed by the Boltzmann factor,
or
(b)~$\sigma$ is turned off because of 
the kinematical threshold of the process.
The Boltzmann suppression is activated 
at $T<M_{H_{\text{SM}}}$ in this case, 
and 
to have $T<M_{H_{\text{SM}}}$ below the blue curve (before hitting the blue curve),
we have to set $\kappa<1.5\cdot 10^{-6}$ which corresponds to $M_{\Delta} \lesssim 670$ GeV with the vevs given in this example.
In other words,
taking into account the LHC bound, the possible range for the triplet mass is narrowed down to $400\text{ GeV} < M_{\Delta} \lesssim 670\text{ GeV}$.
Note that majorons are overproduced from the freeze-in processes with the value of $\kappa$ suggested by this range of $M_{\Delta}$, cf.~Eq.~\eqref{eq:OmegahSq-standard-FI},
if we assume  values of the vevs that can be tested at the XENONnT experiment as in this plot.
The suppression of the process by the kinematical threshold works for $E_{\text{CM}} < M_{\Delta}$, which corresponds to $T < M_{\Delta}/5.4$.
We find that this condition appears on the left of the solid blue curve in \fref{Fig:kappa-equilibrium}, and therefore,
the suppression from the kinematical threshold is not available for any choice of $\kappa$ (hence $M_{\Delta}$).
The other scattering processes induced by the $\kappa$ term provide similar conditions.

In summary, freeze-in production of type-II majorons testable at direct detection experiments has two problems (i) DM overproduction 
and (ii) the condition of non-thermalisation in combination with the LHC bounds.
%
%
However, those two shortcomings can be lifted simultaneously by considering a non-standard cosmic history.
When the BBN starts at $T \sim$ MeV, the universe must be dominated by radiation, however, there are no strict phenomenological requirements on
the development of the universe before that, as we already mentioned in \sref{sec:misalignment}. We can suppose that non-relativistic matter, which is not coupled to radiation thermally, dominates the energy budget of the universe before the radiation-dominated era.
The majoron relic density can be then adjusted by the dilution due to the late-time production of entropy, which is caused by the decay of matter into radiation.
In addition, the Hubble parameter in the early matter-dominated era is modified, and it helps avoid the thermal production of majorons, as we will see later.

\begin{figure}[t!]
\centering
\includegraphics[width=0.52\textwidth]{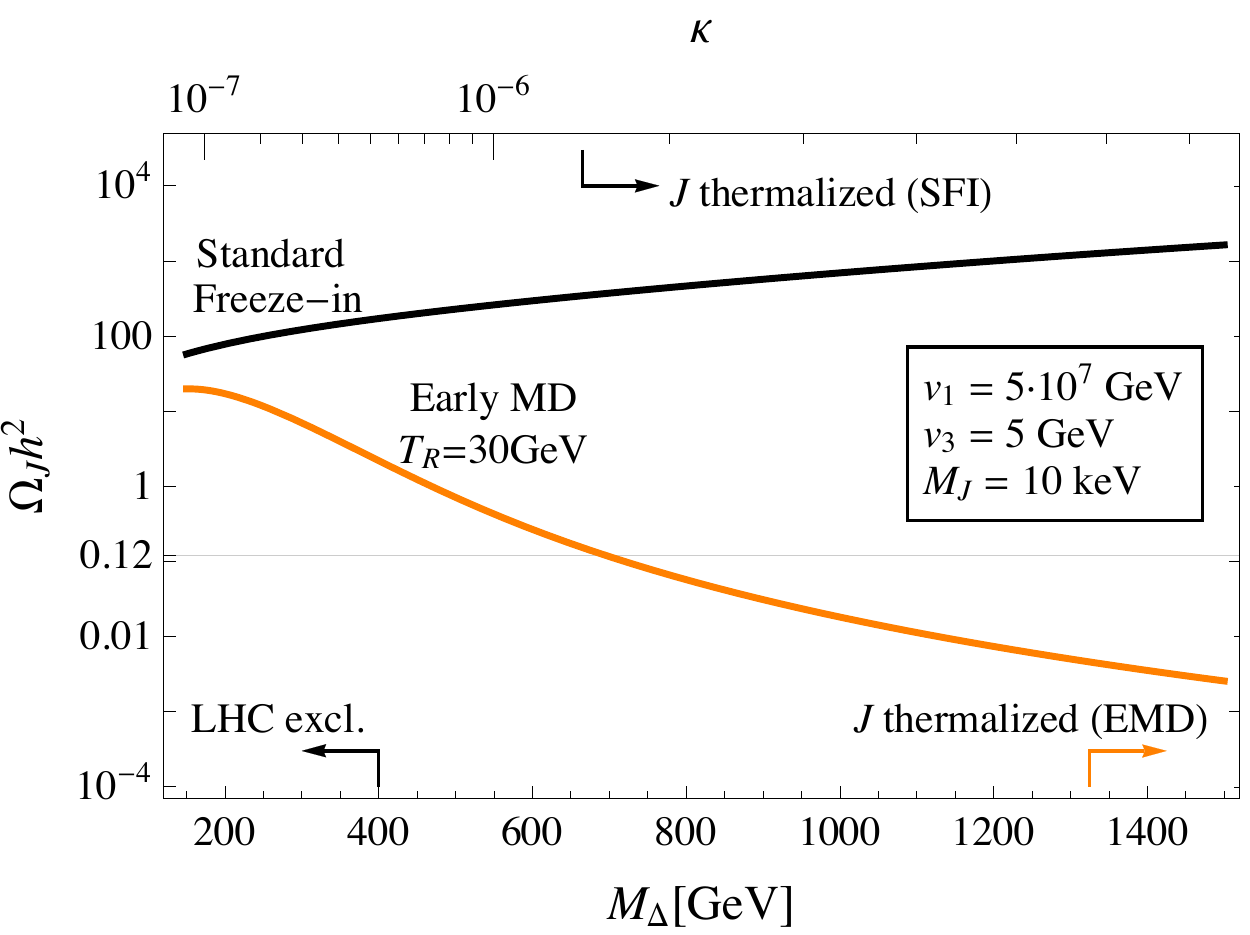}
\hfill
\includegraphics[width=0.44\textwidth]{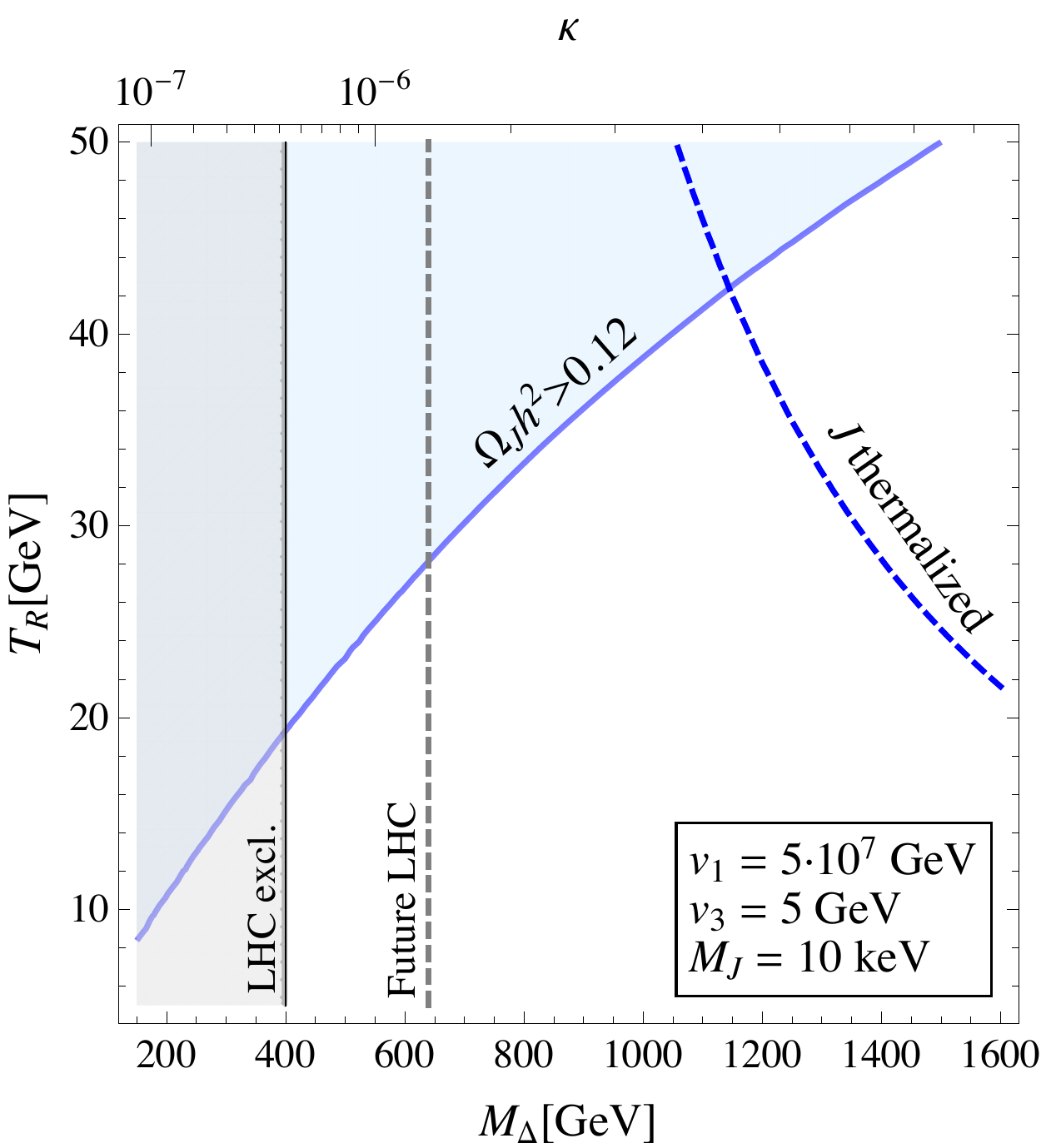}
\caption{[Left] Relic densities of majoron with the standard freeze-in mechanism (black) and 
 the freeze-in in the early matter-dominated era (orange).
 The reheating temperature is set $T_{\text{R}}=30$~GeV
 for the orange curve.
 The bound from extra Higgs boson searches at the LHC is indicated with the label ``LHC excl''.
 In the region indicated as ``$J$ thermalised'', the scalar quartic interactions from the $\kappa$ term thermalise at $T=M_{H_{\text{SM}}}$.
 This bound is denoted by a black arrow for the standard freeze-in case, and by an orange for the freeze-in in the early matter domination.
 [Right]
 Contour on which the correct relic density is reproduced
 in the plane of the triplet scalar mass $M_{\Delta}$ 
 and the reheating temperature $T_{\text{R}}$.
Above the blue dashed curve,
the $\kappa$ quartic interaction comes into equilibrium
and majorons are thermally produced.
}
\label{Fig:typeII-OmegaJhSq}
\end{figure}

Following the discussion in 
Refs.~\cite{Co:2015pka,Calibbi:2021fld}, 
we can calculate the freeze-in production of majorons during the early matter-dominated era with the effect of the dilution due to the late-time entropy injection.
The details of the calculation are presented in Appendix~\ref{App:RelicDensity}.
Here we give an approximate formula
of the relic density from the decay processes:
\begin{align}
\Omega_{J} h^{2}
\simeq
0.12
\left[
\frac{90}{g_{*} (T_{R})}
\right]^{3/2}
\left[
\frac{M_{J}}{10\,\text{keV}}
\right]
\left[
\frac{500\,\text{GeV}}{M_{\Delta}}
\right]^{6}
\left[
\frac{2.7 \cdot 10^{7}\,\text{GeV}}{v_{1}}
\right]^{2}
\left[
\frac{v_{3}}{5\,\text{GeV}}
\right]^{2}
\left[
\frac{T_{R}}{20\,\text{GeV}}
\right]^{7},
\end{align}
which is only valid for $T_{R} \ll M_{\Delta}$.
Setting the reheating temperature 
at $\mathcal{O}(10)$~GeV, the correct relic density can be reproduced with $M_{\Delta} \sim 500$ GeV and the vevs and $M_{J}$ that the XENONnT experiment is sensitive to.
This size of $M_{\Delta}$ together with those reference values of the vevs suggests a relatively large value of $\kappa$, which enhances the rates of the scalar scattering processes leading to thermal production of majorons.
However, in the early matter-dominated era, where $T> T_{R}$, the Hubble parameter is modified as 
\begin{align}
 H (T_{R},T) \simeq
 \sqrt{\frac{4 \pi^{3}}{45} g_{*} (T) }
\frac{T^{4}}{M_{\text{Pl}}  T_{R}^{2}},
\end{align}
cf.~Eqs.~\eqref{eq:Hubble-EMD} and \eqref{eq:rho_RM-as-funcT-EMD} in \aref{App:EarlyMD}.
By setting $T_{R}$ to be smaller than the masses of the scalars, the Hubble parameter is enhanced in comparison with the standard case, and the Gamow's criterion of thermalisation can be fulfilled with a larger value of $\kappa$.
The values of $\kappa$ and $T$ which fulfill $\Gamma = H$ with the reheating temperature $T_{R} = \{50,100\}$~GeV are shown as dashed curves in \fref{Fig:kappa-equilibrium}: for those values of $T_R$
the light blue region would retreat above the dashed lines.

In short, the upper bound to $\kappa$ is significantly relaxed in a scenario with an early matter-dominated era, and we have a chance to reproduce the correct relic density within the parameter region that XENONnT can explore in future. This can be seen in \fref{Fig:typeII-OmegaJhSq}:
in the left panel, the relic density $\Omega_{J} h^{2}$ is shown as a function of $\kappa$\,\,---\,\,and, through \eref{eq:MDelta}, equivalently $M_{\Delta}$\,\,---\,\,as a black solid curve.
Here, the vevs and the mass of the majoron field are fixed at $v_{1} = 5 \cdot 10^{7}$ GeV, $v_{3}=5$ GeV, and $M_{J} = 10$ keV. 
%
The contribution from the scattering processes to the majoron production rate is included, although subdominant, cf. Eq.~\eqref{eq:OmegaJhSq-scattering}.
As discussed above, for standard freeze-in production occurring during a radiation-dominated epoch, the triplet mass is constrained within the range indicated by the black arrows.
The black curve clearly shows that majorons from the standard freeze-in mechanism are overproduced by orders of magnitude for the values of $M_{\Delta}$ in the allowed range.
The orange curve is calculated by assuming an early matter-dominated era with $T_{R}=30$ GeV, which shows that the correct relic density can be reproduced between the LHC bound and the modified bound from the thermalisation condition indicated with the orange arrow.
In the right panel of \fref{Fig:typeII-OmegaJhSq},
we can find the necessary reheating temperature $T_{R}$ in order to reproduce the correct relic density of majorons for a given triplet mass $M_{\Delta}$.
The combination of the relic density condition and the thermalisation condition places the upper bound to the mass of the triplet scalar field, $M_{\Delta} \lesssim 1100$ GeV.
In the future, the LHC bound is expected to reach up to 640 GeV~\cite{Ashanujjaman:2021txz},
and the parameter space will be significantly narrowed down.

Light DM particles produced by the freeze-in mechanism are not thermalised, but a significant amount of them may carry large enough momentum, and this may be in conflict with the bounds on warm DM from structure formation. Lyman-$\alpha$ observations translate into a lower limit on the DM mass in the 10-15~keV range, if standard cosmology is assumed~\cite{Ballesteros:2020adh,DEramo:2020gpr,Decant:2021mhj}.
However, notice that this bound is not directly applicable to the scenario discussed above, which is based on majoron production during an early matter-dominated era. A reconsideration of the Lyman-$\alpha$ bound for such an exotic cosmic history would be needed. For definiteness, in the following, we will regard freeze-in production of majorons with $\mathcal{O}(\text{keV})$ mass as marginally compatible with structure formation.

\section{Constraints on type-II majoron dark matter}
\label{sec:constr}

\subsection{Present status}

Besides the conditions for a substantial production in the early universe discussed in the previous section, a minimal requirement for the majoron to be a viable DM candidate is its stability on cosmological time scales. In other words, the majoron lifetime $\tau_J$ should at least exceed the age of the universe $t_0\approx 13.8$~Gyr. $\tau_J$ can be calculated from the decay rates given in Section~\ref{sec:Jcoupl}.
For a majoron with $M_J\ll m_e$ but still heavier than neutrinos, the decays into the three pairs of neutrino eigenstates dominate. From \eref{eq:Jnunu}, one then gets for the majoron lifetime in this regime:
\begin{align}   
1/\Gamma(J\to \nu\nu) \simeq 15\,\text{Gyr}~\left[\frac{0.01\,\text{eV}^2}{\sum_i m_{\nu_i}^2} \right] \left[\frac{10\,\text{keV}}{M_J} \right] \left[\frac{v_1}{3.8\cdot 10^7\,\text{GeV}} \right]^2 .
\label{eq:taununu}
\end{align}   
As we can see, for a given majoron mass, the requirement $\tau_J > t_0$ translates into a lower bound to $v_1$.
In other words, only a very large lepton-number breaking scale $v_1$ can make the majoron long-lived enough for it to be a DM candidate, which, together with the constraints on $v_2$ and $v_3$ from Eqs.~(\ref{eq:vEW}) and~(\ref{eq:rho}), implies the vev hierarchy anticipated in \eref{eq:vev-hie}.
A somewhat stronger model-independent constraint to the lifetime of majoron DM can be obtained from observations of the cosmic microwave background (CMB), because the late time decay of the DM particles would affect the density fluctuations and the spectrum of the CMB even if the decay products are invisible~\cite{Audren:2014bca,Poulin:2016nat}. Taking into account some tensions with other cosmological data, the most conservative limit is $\tau_{J} > 63$~Gyr~\cite{Poulin:2016nat}.

Unlike decays into neutrinos, majoron decay rates into $\gamma\gamma$ and $e^+e^-$ have a substantial dependence on $v_3$, scaling as $\sim v_3^4$. On the one hand, these modes can be suppressed for small values of $v_3$, since the majoron decouples from fermions in the limit $v_3\to 0$, see \eref{eq:Jcoupl}. On the other hand, if $v_3$ is sizeable, they dominate and entail much stronger constraints on the majoron DM parameter space. For instance, the CMB bound on DM particles decaying into photons or electrons corresponds to lower limits on the partial lifetimes of the order of $10^{24}-10^{25}$~s~\cite{Slatyer:2016qyl}, while from \eref{eq:Jee} we find that our model predicts
\begin{align}   
1/\Gamma(J\to e^+e^-) \simeq 1.2\cdot 10^{25}\,\text{s}~\left[\frac{5\,\text{MeV}}{M_J} \right] \left[\frac{1\,\text{GeV}} {v_3}\right]^4 \left[\frac{v_1}{10^{15}\,\text{GeV}} \right]^2 .
\label{eq:tauee}
 \end{align}   

\begin{figure}[t!]
\centering
\includegraphics[width=1.\textwidth]{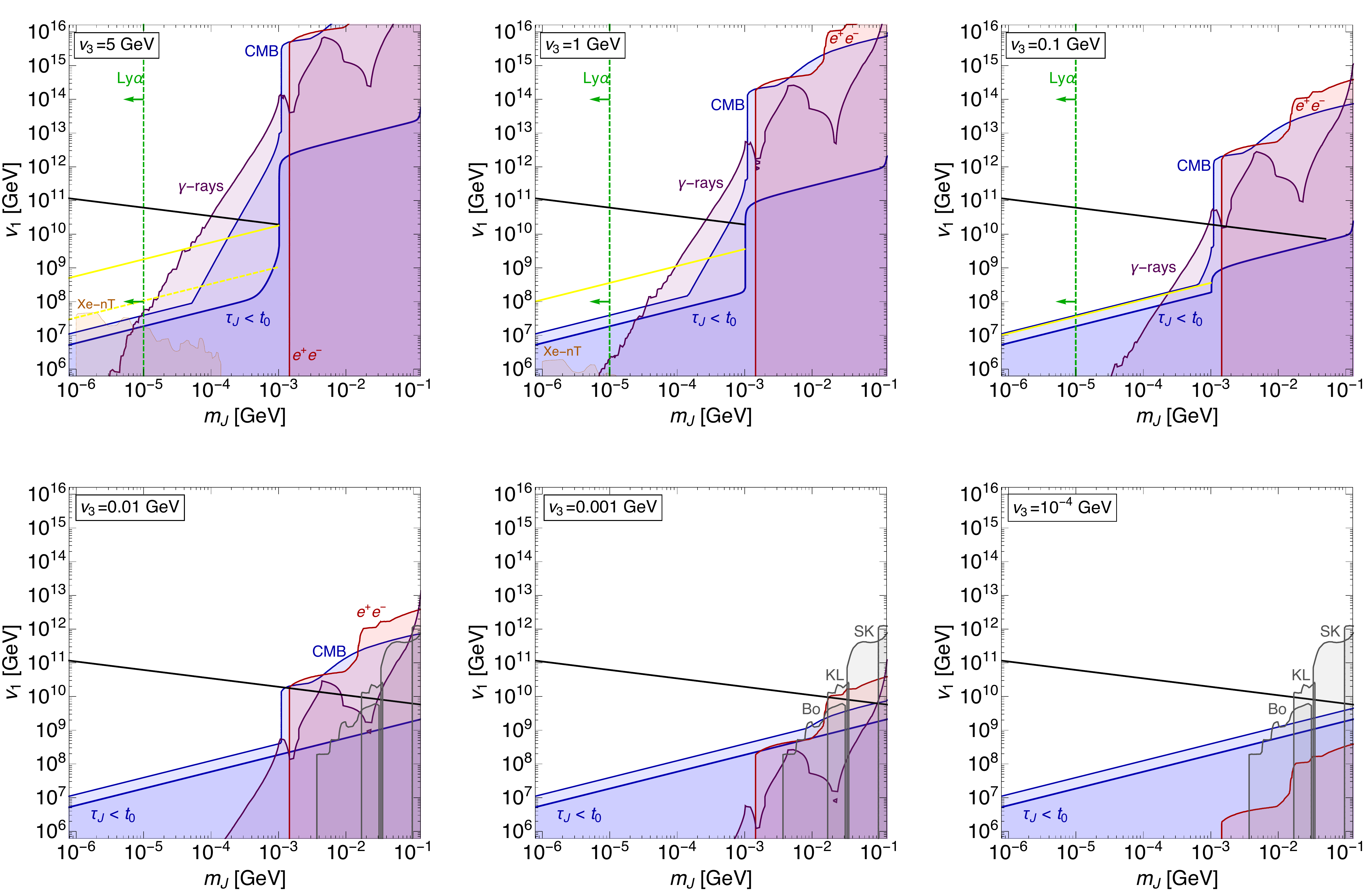}
\caption{Bounds on type-II majoron DM on the $M_J-v_1$ plane for different values of $v_3$. In the dark blue region the majoron lifetime is shorter than the age of the universe (``$\tau_J < t_0$''); the light blue region is excluded by CMB and other cosmological observations (``CMB''); the purple region is excluded by X-ray and gamma-ray data (``$\gamma$-rays''); bounds on majoron DM decaying into electro-positron are shown in red (``$e^+e^-$''); the grey regions are excluded by neutrino experiments: Borexino (``Bo''), KamLAND (``KL''), Super-Kamiokande (``SK''). The orange region represents the direct-detection limit on DM coupling to electrons recently reported by XENONnT ({``Xe-nT''}). The vertical green line is an illustrative lower bound from structure formation on the mass on freeze-in produced majoron DM (``Ly$\alpha$''). The yellow lines correspond to $\Omega_J h^2 = 0.12$ from freeze-in mechanism in a radiation-dominated era (solid) and an early matter dominated-era with $T_R=20$\,GeV (dashed); the black line corresponds to the misalignment mechanism with $\theta_0 =1$. See the text for details and references. 
}
\label{Fig:mJ-v1}
\end{figure}

The above cosmological bounds are summarised in \fref{Fig:mJ-v1} where the regions of the $M_J-v_1$ plane excluded by various searches and constraints are shown for different values of $v_3$: the portion of the plane corresponding to $\tau_J < t_0$ and that excluded by the CMB data~\cite{Poulin:2016nat,Slatyer:2016qyl} are indicated in blue.\footnote{The plots were produced assuming a hierarchical neutrino spectrum, $m_{\nu_1}\ll m_{\nu_2} < m_{\nu_3}$, and normal ordering, that is, $\sum_i m_{\nu_i}^2 \simeq \Delta m^2_\text{atm} + \Delta m^2_\text{sol} \simeq 2.6\cdot 10^{-3}\,\text{eV}^2$, a choice that somewhat increases the majoron lifetime if the neutrino modes are the dominant decays, as one can see from \eref{eq:taununu}.}
Evidently, the CMB constraints are particularly strong for large rates of $J\to \gamma\gamma$ (that require $M_J \sim m_e$ and sizeable $v_3$) and if  $J\to e^+e^-$ is kinematically allowed (and, again, $v_3$ is large enough to provide a substantial coupling of the majoron with electrons), reaching values of $v_1$ as large as $\sim 10^{16}$~GeV.

The oblique lines in \fref{Fig:mJ-v1} indicate where on the $M_J-v_1$ plane the observed DM relic density is achieved, that is, $\Omega_J h^2 = 0.12$, through the two production mechanisms discussed in \sref{sec:DMprod}. The black line corresponds to the misalignment mechanism occurring in the radiation-dominated epoch with $\theta_0 =1$. As we can see from \fref{Fig:misalign}, all of the parameter space above the black line could be compatible with the observed relic density either due to the standard misalignment mechanism or through misalignment during early matter domination.
The yellow lines correspond to the freeze-in mechanism with $M_\Delta = 400$~GeV during a radiation-dominated era (solid line) and an early matter-dominated era with $T_R =20$~GeV (dashed line). In other words, the whole region of the parameter space below the solid yellow line can yield $\Omega_J h^2 = 0.12$ through freeze-in with an appropriate low-scale value of $T_R$. However, as we can see, the freeze-in mechanism ceases to provide sufficient DM production for $v_3\lesssim 0.1$~GeV. 
The vertical green line indicates the approximate lower bound on the mass of freeze-in DM from Lyman-$\alpha$ observations, $M_J \gtrsim 10$~keV~\cite{Ballesteros:2020adh,DEramo:2020gpr,Decant:2021mhj}. As discussed at the end of the previous section, such a constraint can not be straightforwardly applied to freeze-in during an early matter-dominated epoch (dahsed yellow line)\,\,---\,\,although one should not expect that a majoron with $M_J\ll\mathcal{O}(\text{keV})$ be viable. Furthermore, it is well known that mass limits of this kind have no meaning in the case of misalignment-produced ALPs that, being extremely non-relativistic, do not affect structure formation and are thus allowed to be much lighter.

Provided an efficient production mechanism and the fulfillment of the above constraints from cosmological observations, the type-II majoron is a good DM candidate that can be searched for and further constrained by direct-detection and indirect-detection DM experiments. Let us start discussing the latter class of probes. 
Being a decaying DM particle, the type-II majoron will yield signals that, depending on $M_J$, can be searched for by surveys of astrophysical X-rays/gamma-rays, cosmic-ray electrons and positrons, and neutrino lines. 

Stringent constraints on light DM decaying into photons can be obtained from the measurements of the diffuse X-ray and gamma-ray galactic and extra-galactic backgrounds performed by a variety of satellite telescopes. For the range of DM masses that we are considering here, these bounds translate into lower limits on the partial lifetime $\tau_{\gamma\gamma}\equiv 1/\Gamma(J\to \gamma\gamma)$ in the range $10^{26}-10^{29}$~s. The purple-shaded regions in \fref{Fig:mJ-v1} show the the impact of these constraints on the type-II majoron parameter space. For $M_J$ up to $\mathcal{O}$(MeV) we employed the limits on $\tau_{\gamma\gamma}$ collected in Ref.~\cite{Panci:2022wlc}, including the bound from the INTEGRAL satellite as recently reevaluated in Ref.~\cite{Laha:2020ivk}. For heavier majorons, we adopted the results of Ref.~\cite{Essig:2013goa}.

Above the kinematic threshold $M_J = 2m_e$, a distinctive signature of majoron DM stems from the decay $J\to e^+e^-$. The red regions in \fref{Fig:mJ-v1} are excluded by constraints on DM decaying into $e^+e^-$ as obtained in Ref.~\cite{Boudaud:2016mos} considering the cosmic-ray data of the Voyager satellite, and in Ref.~\cite{Cirelli:2023tnx} analysing the X-ray data collected by XMM-Newton. This latter bound stems from the upscattering of low-energy galactic photons due to the electron pair that would generate an X-ray emission. We find that it is more stringent than the Voyager limit with the exception of the mass window $16\,\text{MeV}\lesssim M_J \lesssim 40\,\text{MeV}$.
 
Constraints on DM decaying into neutrinos can be obtained from the fluxes of cosmic neutrinos~\cite{Palomares-Ruiz:2007egs} measured by experiments such as Borexino~\cite{Borexino:2010zht}, KamLAND~\cite{KamLAND:2011bnd}, Super-Kamiokande~\cite{Super-Kamiokande:2002exp,Super-Kamiokande:2013ufi}. Here, we employ the bounds that these experiments set on potential neutrino lines as evaluated in Ref.~\cite{Garcia-Cely:2017oco}\,\,---\,\,see also Refs.~\cite{Coy:2020wxp,Arguelles:2022nbl,Akita:2023qiz}. 
Since the majoron field directly couples to neutrino mass eigenstates, the neutrinos produced in majoron decays do not undergo oscillations~\cite{Garcia-Cely:2017oco}.  Consequently, the flavour composition of the neutrino flux is the same at the source and at the detection point, and is given by
\begin{align}        
\label{eq:alphas}
\alpha_\ell = \frac{\sum_i m_{\nu_i}^2|U_{\ell i}|^2}{\sum_i m_{\nu_i}^2}
\end{align}
(where $U_{\ell i}$ are entries of the PMNS matrix and $\ell = e,\,\mu,\,\tau$) such that the neutrino and antineutrino fluxes for the flavour $\ell$ are $\propto \alpha_\ell \,\Gamma(J\to \nu\nu)$~\cite{Garcia-Cely:2017oco}. Hence the quantities $\alpha_\ell$ control the relative importance of searches targeting different neutrino flavours and how they compare with the CMB bound that depends only on $\Gamma(J\to \nu\nu)$.
In \fref{Fig:mJ-v1}, in order to compare the type-II majoron decay rates with the limits reported in Refs.~\cite{Garcia-Cely:2017oco,Coy:2020wxp} (with the Super-Kamiokande limit adapted from Ref.~\cite{Palomares-Ruiz:2007egs}), we assume normal ordering, $m_{\nu_1}\ll m_{\nu_2} < m_{\nu_3}$, and take the best-fit values of the oscillation parameters from Ref.~\cite{Esteban:2020cvm}.
The impact of other possible choices for the neutrino parameters will be discussed in the next subsection. The resulting excluded regions are depicted in grey.
Notice that, for values of $v_3 \lesssim 10^{-3}$~GeV, the only searches that are currently sensitive to our majoron DM parameter space are those performed employing neutrino data. In other words, for small values of $v_3$, the majoron DM phenomenology is dominated by the $J\to \nu\nu$ decays that do not depend on $v_3$ itself,\footnote{This follows from the coupling with neutrinos shown in \eref{eq:g_Jphph-typeII-in-app} in the limit of \eref{eq:vev-hie} that, as we argued, is necessary for majorons of cosmological interest.} hence the bounds shown in the last panel of the figure apply to lower values of $v_3$ as well. 
Future prospects of all kinds of searches for decaying DM are discussed in the next subsection.

The electron-majoron interaction can also give rise to an electron recoil signal detectable at DM direct-detection experiments.\footnote{For ALP explanations, including related tests and possible issues, of the electron recoil signal reported by XENON1T~\cite{XENON:2020rca} but later excluded by XENONnT~\cite{XENON:2022ltv}, cf.~\emph{e.g.}~Refs.~\cite{Takahashi:2020bpq,Bloch:2020uzh,DiLuzio:2020jjp,Arias-Aragon:2020qtn,Han:2020dwo,Takahashi:2020uio,Buttazzo:2020vfs,Han:2022iig,Sakurai:2022roq}.} 
The currently most sensitive search has been performed by XENONnT with an exposure of $1.16$~ton$\times$year and interpreted in terms of ALP DM with mass in the range $1-130$~keV, yielding upper limits on the ALP coupling with electrons of the order $10^{-14}-10^{-13}$~\cite{XENON:2022ltv}. In our case, from \eref{eq:Jcoupl}, we get for the majoron-electron coupling 
\begin{align}      
\left|g_{J e e}^{P}\right| \simeq \frac{2 m_e v_3^2}{v_1v_2^2}
\simeq 1.7 \cdot 10^{-15}\, \left[\frac{10^7\,\text{GeV}}{v_1}\right]
\left[\frac{v_3}{1\,\text{GeV}}\right]^2\,.
\end{align}    
Hence, XENONnT is sensitive to keV-scale majoron DM evading the CMB bound\,\,---\,\,which requires $v_1 > 10^7$~GeV, as we have seen above\,\,---\,\,provided that $v_3 > 1$\,GeV, a value close to the current upper limit from electro-weak precision observables, $v_3 < 7$~GeV.
In \fref{Fig:mJ-v1}, the region excluded by XENONnT is shown in orange. As discussed in the previous section, the corner of the parameter space that direct-detection experiments are sensitive to can be compatible with the observed DM relic density only if our type-II majoron is produced through the freeze-in mechanism occurring during an early matter-dominated era (yellow dashed line).
It is well known that a keV-scale (or lighter) long-lived ALP emitted by electrons in stars would provide an extra cooling process affecting stellar evolution~\cite{Raffelt:1990yz}. On the other hand, present astrophysical observations are largely compatible with the standard cooling process due to the emission of neutrinos and disfavour ALP coupling with electrons larger than $\mathcal{O}(10^{-13})$~\cite{Raffelt:1994ry,Viaux:2013lha,MillerBertolami:2014rka}. Hence star cooling bounds are currently weaker than the limit provided by XENONnT in the region of the parameter space we are considering.

\begin{figure}[t!]
\centering
\includegraphics[width=0.9\textwidth]{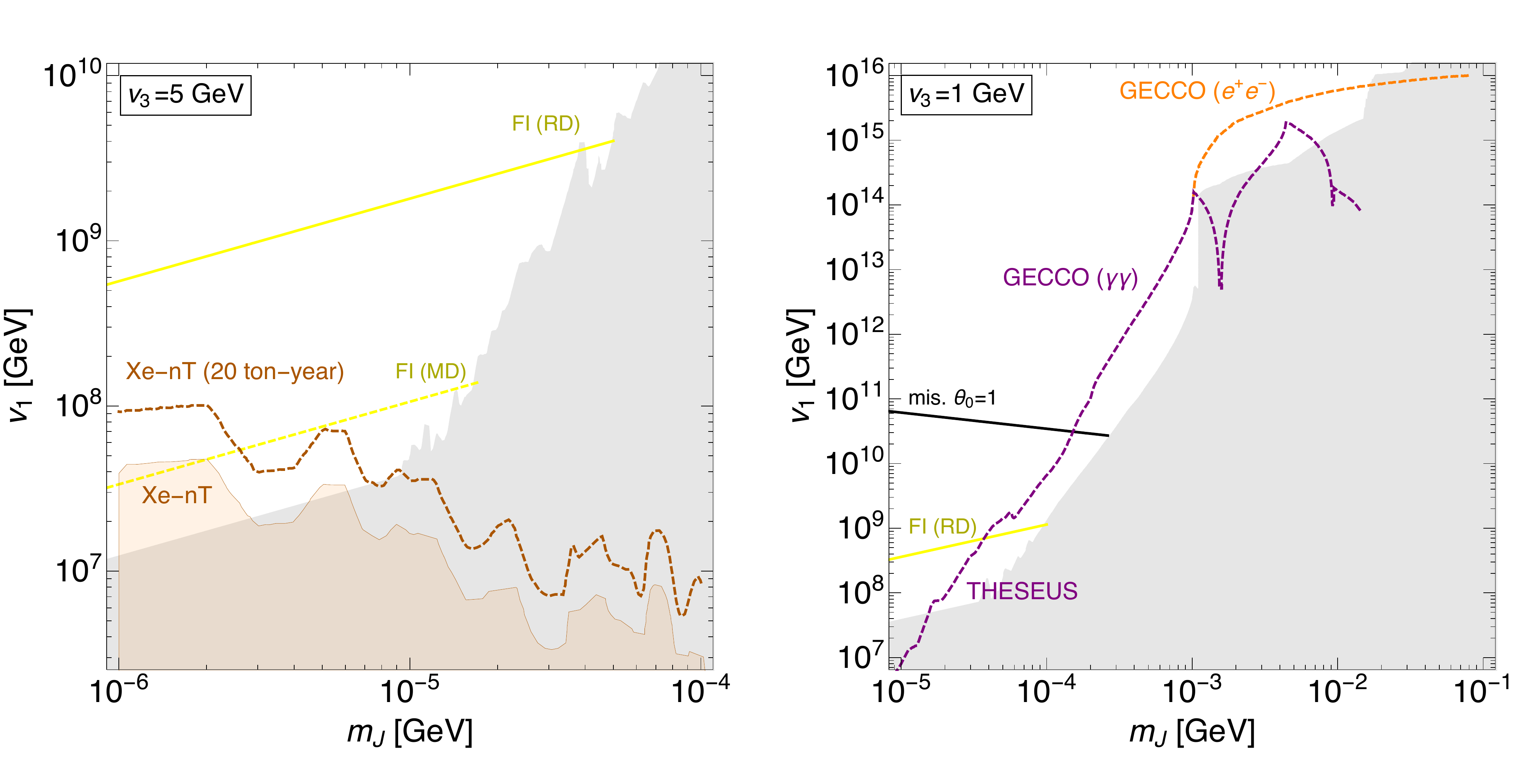}
\caption{
Prospected sensitivity to the type-II majoron of direct (left plot) and indirect (right plot) DM searches for two illustrative choices of $v_3$. 
The current XENONnT bound is shown as an orange region, while  the gray-shaded areas are excluded by the other constraints displayed in \fref{Fig:mJ-v1}.
The dark orange dashed line depicts the sensitivity of XENONnT with the planned exposure of 20 ton$\times$year. 
The purple (orange) dashed line shows the impact of future constraints on DM decaying into $\gamma\gamma$ ($e^+e^-$). The oblique lines correspond to $\Omega_J h^2 = 0.12$ as in \fref{Fig:mJ-v1}. See the text for further details.
}
\label{Fig:fut}
\end{figure}

\subsection{Future prospects}

Let us now move to consider the prospect of testing type-II majoron DM at future experiments. 
As we have seen above, searches for DM-electron scattering at direct detection experiments recently started to test a corner of our majoron paremeter space, for $M_J \sim 1-10$~keV and large values of $v_3$, that is, close to the EWPO limit. In the left plot of \fref{Fig:fut}, we show the projected sensitivity of XENONnT as obtained by naively rescaling the expected limit reported in Ref.~\cite{XENON:2022ltv} according to the final exposure goal of the experiment\,\,---\,\,20 ton$\times$year~\cite{XENON:2020kmp}. A comparable exposure is to be expected at analogous direct-detection experiments such as PandaX~\cite{PandaX:2022ood} and  LZ~\cite{Mount:2017qzi}.
As we can see, the limit on $v_1$ will be improved by roughly a factor of two, up to $v_1 \sim 10^8$~GeV.
That direct detection experiments can be to any extent sensitive to this instance of ALP DM is a remarkable consequence of the fact that the lepton number is anomaly free, hence the majoron decouples from photons for $M_J\ll m_e$, see \eref{eq:Jgammas}. Otherwise, this corner of the parameter space would be completely excluded by X-ray constraints, as discussed in the literature addressing the (now excluded) excess observed in the XENON1T experiment~\cite{XENON:2020rca}\,\,---\,\,see, in particular, Refs.~\cite{Takahashi:2020bpq,Bloch:2020uzh}.

\begin{figure}[t!]
\centering
\includegraphics[width=0.9\textwidth]{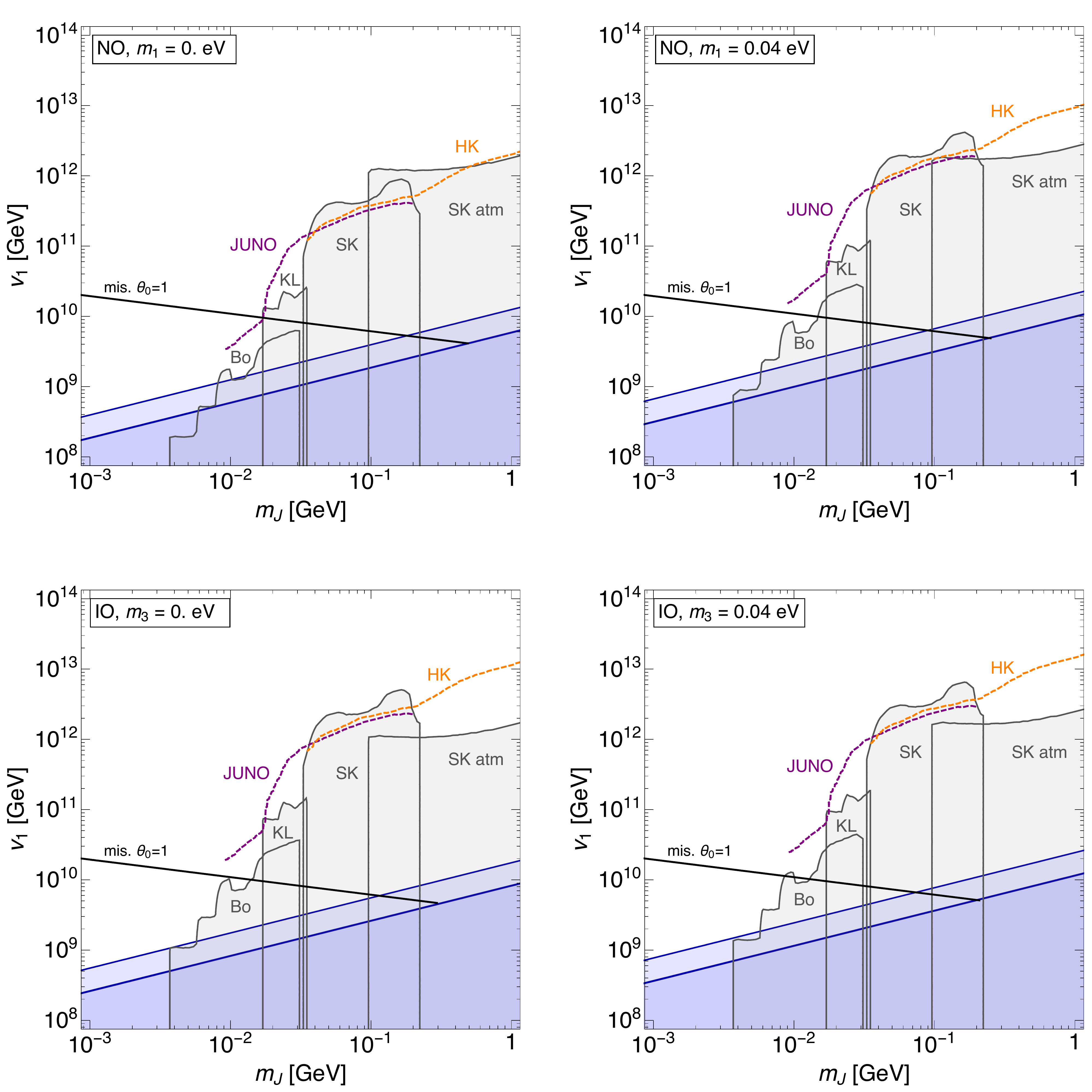}
\caption{Current and future expected limits on the type-II majoron DM parameter space from neutrino experiments for different choices of the neutrino mass spectrum. First line: normal mass ordering (NO); Second line: inverted mass ordering (IO). The excluded blue regions and the $\Omega_J h^2 =0.12$ line from misalignment are as in \fref{Fig:mJ-v1}. Present limits are shown in gray, future limits from JUNO and Hyper-Kamiokande~(``HK'') are indicated by dashed purple lines and orange lines respectively. See the text for details.
}
\label{Fig:fut-nut}
\end{figure}

In the right panel of \fref{Fig:fut} we show the future constraints on decaying DM for $v_3 = 1$~GeV. Decreasing the value of $v_3$, the limit on $v_1$ just scales as $\sim v_3^2$, until  sensitivity is lost for $v_3\lesssim 10^{-3}$~GeV, as shown in \fref{Fig:mJ-v1}. The purple line shows the reach of future searches for DM decaying into photons. For small $M_J$ the most sensitive probe will be provided by the instruments aboard the THESEUS mission~\cite{Thorpe-Morgan:2020rwc}, while the GECCO telescope~\cite{Coogan:2021rez} is expected to give the best prospects for $M_J \gtrsim 0.2$~MeV. For a compilation of the future limits on decaying DM in the X-ray to soft gamma-ray region, see Ref.~\cite{Panci:2022wlc}. GECCO will also be sensitive to DM decaying into electrons~\cite{Coogan:2021rez}. As the orange dashed line shows, this can yield a factor of ten improvement on the bound on $v_1$ for $1\,\text{MeV}\lesssim M_J \lesssim 10\,\text{MeV}$.

The prospects of future neutrino experiments\,\,---\,\,adapted from the results of Ref.~\cite{Arguelles:2022nbl}\,\,---\,\,are displayed in \fref{Fig:fut-nut} by dashed lines: orange for Hyper-Kamiokande (HK)~\cite{Hyper-Kamiokande:2018ofw}, purple for JUNO~\cite{JUNO:2015zny}. As in \fref{Fig:mJ-v1}, the lifetime/CMB bounds and the current neutrino constraints are shown in blue and gray respectively. 
For the former constraints, we take $\tau_J = 1/\Gamma(J\to\nu\nu)$, that is, they are valid as long as $v_3 \lesssim 10^{-3}$~GeV, as we can see from \fref{Fig:mJ-v1}, while the bounds set by neutrino experiments do not depend on $v_3$. As we mentioned, for small enough values of $v_3$, these are the only relevant limits on the parameter space of type-II majoron DM. In \fref{Fig:fut-nut} we show values of $M_J$ up to 1~GeV, hence, compared to \fref{Fig:mJ-v1}, we can additionally display a limit from the measurement of the atmospheric muon neutrino flux at Super-Kamiokande (``SK atm'') as derived in Ref.~\cite{Palomares-Ruiz:2007egs}.\footnote{Notice that $M_J = 1$~GeV is above the kinematic thresholds of majoron decays into muons and hadrons, $J\to \mu^+\mu^-$ and $J\to \pi\pi\pi$. Nevertheless, we checked that the resulting constraints are still subdominant as long as $v_3 \lesssim 10^{-4}$~GeV.} 
The relative importance of this search compared to the other limits from neutrino experiments (based on observations of $\bar \nu_e$) and the bounds on the majoron lifetime depends on the flavour composition of the the neutrino flux given by \eref{eq:alphas}.
In order to gauge the dependence of $\alpha_\ell$ on the neutrino parameters, we show in \fref{Fig:fut-nut} four different possible neutrino mass spectra, still setting the oscillation parameters to the best-fit values reported in Ref.~\cite{Esteban:2020cvm}: normal mass ordering (NO) with $m_{\nu_1} = 0$ or $0.04$~eV; inverted mass ordering (IO) with $m_{\nu_3} = 0$ or $0.04$~eV.\footnote{We remind that for NO, the spectrum is such that $m_{\nu_1} < m_{\nu_2}\,\left[=\sqrt{m_{\nu_1}^2+\Delta m^2_\text{sol}}\right] < m_{\nu_3}\,\left[=\sqrt{m_{\nu_1}^2+\Delta m^2_\text{atm}}\right],$ while for IO one has 
$m_{\nu_3} < m_{\nu_1}\,\left[=\sqrt{m_{\nu_3}^2+\Delta m^2_\text{atm}}\right] < m_{\nu_2}\,\left[=\sqrt{m_{\nu_1}^2+\Delta m^2_\text{sol}}\right],$ where $\Delta m^2_\text{atm}$ and $\Delta m^2_\text{sol}$ are respectively the atmospheric and solar neutrino mass splittings.}
Notice that the largest value of the absolute mass scale is chosen so as to be (marginally) compatible with the cosmological upper limit $\sum_i m_{\nu_i} < 0.13$~eV~\cite{DES:2021wwk}.
The above choices for the neutrino mass spectrum yield
\begin{align*}
\text{NO:}\quad (\alpha_e,\alpha_\mu,\alpha_\tau) = (0.03,0.55,0.42)~[m_{\nu_1}=0],~~(0.23,0.41,0.36)~[m_{\nu_1}=0.4\,\text{eV}]\,. \\
\text{IO:}\quad (\alpha_e,\alpha_\mu,\alpha_\tau) = (0.49,0.22,0.29)~[m_{\nu_3}=0], ~~(0.41,0.28,0.31)~[m_{\nu_3}=0.4\,\text{eV}]\,. 
\end{align*}
The accidental suppression of the electron portion of the neutrino flux for NO with $m_{\nu_1}=0$ explains why, in the top-left panel of \fref{Fig:fut-nut}, the Borexino limit barely extends beyond the CMB limit and, in the large mass regime, the future HK sensitivity does not seem to be able to improve over the limit from atmospheric neutrinos. As we can see from the other plots, this is not the case for the other three considered spectra, for which, in particular, Hyper-Kamiokande may improve the current bound by about one order of magnitude.

Let us recall that, above the black lines in Figures~\ref{Fig:fut} and \ref{Fig:fut-nut}, the whole parameter space can be compatible with the observed DM relic abundance through the misalignment mechanism.
These figures thus show that depending on $M_J$, $v_3$, and the details of the production mechanism, all of the three considered decay modes ($\gamma\gamma$, $e^+e^-$, $\nu\nu$) of majoron DM particles could yield observable signals at upcoming experiments.

\section{Summary and conclusions}
\label{sec:concl}

In this study, we have worked within the context of type-II seesaw, which is perhaps the most economical model to address the origin of neutrino masses, one of the outstanding questions in particle physics.
Besides providing a simple UV completion to the neutrino Majorana mass terms, type-II seesaw enjoys other theoretically and phenomenologically desirable features. To name a few, the triplet scalar in type-II seesaw can account for the observed baryon asymmetry through leptogenesis~\cite{Barrie:2021mwi,Barrie:2022cub} and it can also play a fundamental role in gauge coupling unification within the context of minimal grand unified theories~\cite{Dorsner:2005fq,Dorsner:2005ii,Dorsner:2006dj,Dorsner:2006hw,Dorsner:2007fy,FileviezPerez:2008dw,Antusch:2022afk,Calibbi:2022wko}.

Here, we considered a minimal extension of the type-II seesaw mechanism that dynamically addresses the breaking of the lepton number by introducing an additional scalar singlet~\cite{Choi:1989hj,Choi:1991aa}.
We showed that the resulting PNGB, the type-II majoron, is an excellent dark matter candidate, thus adding the nature of dark matter to the number of outstanding problems that type-II seesaw can account for. We performed the
first systematic study of the production of type-II majoron DM in the early universe and its possible signals at direct and indirect detection experiments. The latter searches can be sensitive to our model, because the type-II majoron is an instance of DM that, depending on its mass, decays into $e^+e^-$, $\gamma\gamma$, $\nu\nu$.

We have shown that type-II majorons can account for the measured DM relic abundance in its entirety, if produced through either the freeze-in mechanism or the misalignment mechanism.
Freeze-in production can occur through the decay of the heavy states belonging to the scalar triplet into a majoron and a SM Higgs or gauge boson, see \eref{eq:FI-decay-process}, which 
requires a triplet mass $M_\Delta \lesssim 1$~TeV, hence light enough to be tested at the LHC and/or future colliders. Provided that, freeze-in production can be effective, while the model can evade at the same time constraints on decaying dark matter, up to values of the lepton-number breaking vev $v_1$ of the order of $10^9$~GeV if the triplet vev $v_3$ is $\mathcal{O}(1)$~GeV, see \fref{Fig:mJ-v1}. For lower values of $v_1$, majoron relic density can be made consistent with cosmological observations either by decreasing $v_3$ or by taking a low value of the reaheating temperature $T_R$, which would imply freeze-in production taking place during an early matter dominated era so causing a dilution of the final relic abundance, as extensively discussed in \sref{sec:DMprod}. In any case, \fref{Fig:mJ-v1} shows that, below $v_3 \simeq 0.1$~GeV, freeze-in production ceases to be viable in particular due to CMB constraints. This latter bound, in combination with lower limits on the DM mass from structure formation, also implies that majoron DM requires $v_1 \gtrsim 10^7$~GeV.
Provided that the above conditions are fulfilled, the freeze-in mechanism can effectively produce type-II majoron DM for majoron masses in the range $1-100$~keV.

Misalignment production of type-II majorons can account for the entirety of the observed DM relic density while being compatible with bounds on decaying DM for $v_1 \gtrsim 10^{10}$~GeV. Below that value, majorons are always a subdominant DM component (unless another production mechanism is at work, such as freeze-in), while above it the majoron relic abundance can match the observed one if one decreases the value of the initial misalignment angle $\theta_0$ or, again, if DM production occurs during an early matter dominated era, which requires a low value of the reheating temperature, cf.~\sref{sec:DMprod} and  \fref{Fig:misalign}.

In \sref{sec:constr}, we have extensively discussed the constraints on and discovery prospects of the type-II majoron in the regime where it is a viable DM candidate, as following from its decay modes and coupling with electrons. For small values of the triplet vev, $v_3 \lesssim 10^{-3}$~GeV, the majoron phenomenology is dominated by its couplings with neutrinos, because those with other fermions (and consequently photons) are suppressed by a factor $\sim v_3^2 / v_2^2 \simeq v_3^2/v^2_\textrm{EW}$. In this regime, type-II majoron DM is subject to constraints from neutrino experiments and its decay $J\to \nu\nu$ can be a target for searches of monochromatic neutrino lines at upcoming neutrino telescopes as long as $M_J \gtrsim 10$~MeV\,\,---\,\,see \fref{Fig:fut-nut}. If $v_3 > 10^{-3}$~GeV, the decays $J\to e^+e^-$ and $J\to \gamma\gamma$ can give rise to observable signals at future X-ray and soft gamma-ray probes such as GECCO~\cite{Coogan:2021rez}, for values of $M_J$ as low as $M_J \approx 10$~keV, see Figures~\ref{Fig:mJ-v1} and \ref{Fig:fut} (right).
For lower majoron masses, we found a corner of the parameter space\,\,---\,\, $v_1 \approx 10^7-10^8$~GeV, $1\,\text{keV}\lesssim M_J \lesssim 10\,\text{keV}$, $v_3 > 1$~GeV, a regime suitable for freeze-in DM production\,\,---\,\,where the type-II majoron can give an electron recoil signal observable at direct detection experiments such as XENONnT~\cite{XENON:2022ltv}, see \fref{Fig:fut} (left). This is a consequence of the fact that, being the lepton number free of electromagnetic anomalies, majorons  enjoy suppressed coupling with photons for $M_J\ll m_e$, which makes them a plausible target for direct detection experiments in a regime where other ALP DM candidates are excluded by X-ray data~\cite{Takahashi:2020bpq,Bloch:2020uzh}.

Finally, another distinctive feature of the type-II majoron that we found is that its interactions with SM fermions are  \emph{flavour conserving}, being inherited from mixing with the Higgs doublet. This gives rise to a different phenomenology compared to other ALPs of cosmological interest (including the majorons from type-I seesaw~\cite{Heeck:2019guh} and type-III seesaw~\cite{Cheng:2020rla}) that are instead tightly constrained by searches for two-body flavour-violating decays of mesons or leptons into an invisible ALP $a$, such as $K\to \pi a$ and $\mu\to e a$, see Refs.~\cite{Calibbi:2016hwq,MartinCamalich:2020dfe,Calibbi:2020jvd,Jho:2022snj,Panci:2022wlc,DEramo:2021usm,DiLuzio:2023ndz}. Hence flavour processes of such kind are not only a promising avenue to search for a wide class of ALP DM candidates but they can also provide a handle for model discrimination in case of positive signals.


\medskip\paragraph{Acknowledgements.} 
It is a pleasure to thank Enrique Fern\'{a}ndez-Mart\'{i}nez, Luca Merlo, Sergio Palomares-Ruiz, Mathias Pierre for very valuable discussions. LC acknowledges support from the National Natural Science Foundation of China under the grants No.~12035008 and No.~12211530479.
TO acknowledges partial financial support by the Spanish Research Agency (Agencia Estatal de Investigaci\'{o}n) through the grant IFT Centro de Excelencia Severo Ochoa No CEX2020-001007-S and by the grant PID2019-108892RB-I00 funded by MCIN/AEI/10.13039/501100011033. TO also acknowledges support through the European Union's Horizon 2020 research and innovation programme under the Marie Sklodowska-Curie grant agreements No 860881-HIDDeN and No 101086085-ASYMMETRY.

\appendix

\section{Details on the type-II majoron model}
\label{App:typeII}

In the discussion of the model, we mainly follow the notation adopted in Ref.~\cite{Bonilla:2015jdf}.
As discussed in \sref{sec:typeII}, in addition to the $SU(2)_{L}$ doublet scalar $\Phi$,
which is the SM Higgs field,
singlet and triplet scalars, $\sigma$ and $\Delta$, are 
introduced,\footnote{%
%
Notice that the triplet field $\Delta$ is defined differently from the literature on the triplet Higgs model such as Refs.~\cite{Aoki:2011pz,Aoki:2012jj,Kanemura:2013vxa,
Kanemura:2014goa,Ashanujjaman:2021txz} 
by an anti-symmetric tensor (and also some signs).
}
\begin{gather}
\sigma
=
\frac{1}{\sqrt{2}}
\left(v_{1} + R_{1} + {\rm i} I_{1} \right),
\quad 
\Phi = 
\begin{pmatrix}
\frac{1}{\sqrt{2}}
\left(v_{2} + R_{2} + {\rm i} I_{2} \right)
\\
\phi^{-}
\end{pmatrix},
\\
\Delta=
\begin{pmatrix}
\frac{1}{\sqrt{2}}
\left(v_{3} + R_{3} + {\rm i} I_{3} \right)
& \Delta^{+}/\sqrt{2}
\\
\Delta^{+}/\sqrt{2} & \Delta^{++}
\end{pmatrix},
\end{gather}
where $v_{1,2,3}$ are the vacuum expectation values,
$R_{1,2,3}$ and $I_{1,2,3}$ are the real and imaginary part 
of the neutral components.
The triplet scalar carries lepton number $-2$
so that it can have a Yukawa interaction with 
two lepton doublets, cf.~Eq.~\eqref{eq:L_Yukawa-typeII}.
The doublet scalar, which is to be identified with the SM Higgs scalar, 
does not have lepton number charge.
The lepton number assignment of the singlet scalar 
is $+2$ so that 
the $\sigma \Phi^{\sf T} \Delta \Phi$ term becomes
invariant under the $U(1)$ lepton number transformation.
In this setup, the lepton number is spontaneously broken by 
the vacuum expectation value of the singlet field, which 
provides a Nambu-Goldstone boson, the majoron, in the imaginary 
component of the singlet field.\footnote{%
We assume that there is a small explicit lepton number 
violating term in the scalar potential, which provides 
a mass to the majoron field, although we will not write 
the term explicitly in Eq.~\eqref{eq:typeII-scalar-potential}.
}

Invariance under the SM gauge symmetries and the lepton number determine
the scalar potential as follows
\begin{align}
-\mathscr{L}_{S} 
=&
\mu_{1}^{2} \sigma^{*} \sigma
+
\mu_{2}^{2}
\Phi^{\dagger} \Phi 
+
\mu_{3}^{2}
\text{Tr}
\left[
\Delta^{\dagger} \Delta
\right]
+
\lambda_{1}
\left(
\Phi^{\dagger} \Phi 
\right)^{2}
+
\lambda_{2} 
\left(
\text{Tr}
\left[
\Delta^{\dagger} \Delta
\right]
\right)^{2}
\nonumber
\\
&
+
\lambda_{3}
\Phi^{\dagger} \Phi
\text{Tr}
\left[
\Delta^{\dagger} \Delta
\right]
+
\lambda_{4}
\text{Tr}
\left[
\Delta^{\dagger} \Delta
\Delta^{\dagger} \Delta
\right]
+
\lambda_{5}
\text{Tr}
\left[
\Phi^{\dagger} \Delta^{\dagger}
\Delta \Phi
\right]
\nonumber
\\
&
+
\beta_{1}
\left( \sigma^{*} \sigma \right)^{2}
+
\beta_{2}
\left( \sigma^{*} \sigma \right)
\left( \Phi^{\dagger} \Phi \right)
+
\beta_{3}
\left( \sigma^{*} \sigma \right)
\text{Tr}
\left[
\Delta^{\dagger} \Delta
\right]
-
\kappa
\left[ 
\sigma
\Phi^{\sf T} \Delta \Phi 
+
\text{H.c.}
\right].
\label{eq:typeII-scalar-potential}
\end{align}
The mass matrices of the component fields
follow from the quadratic terms in the potential: 
\begin{align}
-\mathscr{L}_{S} 
\supset
\frac{1}{2}
R_{a} \left(M_{R}^{2}\right)_{ab} R_{b}
+
\frac{1}{2}
I_{a} \left(M_{I}^{2}\right)_{ab} I_{b}
+
S^{-}_{a}
(M_{H^{\pm}}^{2})_{ab}
S^{+}_{b}
+
\Delta^{--}
M_{H^{\pm\pm}}^{2}
\Delta^{++},
\end{align}
where the singly-charged scalars are labelled as
\begin{align}
S^{\pm}_{1} = \phi^{\pm}, \quad S^{\pm}_{2} = \Delta^{\pm},
\end{align}
and the mass matrices are 
\begin{align}
(M_{R}^{2})_{ab}
=&
\begin{pmatrix}
2 \beta_{1} v_{1}^{2} + \frac{1}{2} \kappa \frac{v_{2}^{2} v_{3}}{v_{1}}
&
\beta_{2} v_{1} v_{2} - \kappa v_{2} v_{3}
&
\beta_{3} v_{1} v_{3} - \frac{1}{2} \kappa v_{2}^{2}
\\
\beta_{2} v_{1} v_{2} - \kappa v_{2} v_{3}
&
2 \lambda_{1} v_{2}^{2}
&
(\lambda_{3} + \lambda_{5}) v_{2} v_{3} - \kappa v_{1} v_{2}
\\
\beta_{3} v_{1} v_{3} - \frac{1}{2} \kappa v_{2}^{2}
&
(\lambda_{3} + \lambda_{5}) v_{2} v_{3}
-
\kappa v_{1} v_{2}
&
2 (\lambda_{2} + \lambda_{4}) v_{3}^{2}
+
\frac{1}{2} \kappa \frac{v_{1} v_{2}^{2}}{v_{3}}
\end{pmatrix},
\\
(M_{I}^{2})_{ab}
=&
\kappa
\begin{pmatrix}
\frac{1}{2} \frac{v_{2}^{2} v_{3} }{v_{1}}
&
v_{2} v_{3}
&
\frac{1}{2} v_{2}^{2}
\\
v_{2} v_{3} 
&
2 v_{1} v_{3}
&
v_{1} v_{2}
\\
\frac{1}{2} v_{2}^{2}
&
v_{1} v_{2}
&
\frac{1}{2} \frac{v_{1} v_{2}^{2}}{v_{3}}
\end{pmatrix},
\\
(M_{H^{\pm}}^{2} )_{ab}
=&
\left(\kappa v_{1} - \frac{1}{2} \lambda_{5} v_{3} \right)
\begin{pmatrix}
v_{3}
&
-
\frac{1}{\sqrt{2}} 
v_{2} 
\\
-
\frac{1}{\sqrt{2}} 
v_{2} 
&
\frac{1}{2}
\frac{v_{2}^{2}}{v_{3}}
\end{pmatrix},
\\
M_{H^{\pm \pm}}^{2}
=&
\frac{1}{2} 
\left(
\kappa 
\frac{v_{1} v_{2}^{2}}{v_{3}}
-
2 
\lambda_{4} v_{3}^{2}
-
\lambda_{5} v_{2}^{2}
\right).
\end{align}
The mass eigenstates,
$H_{i}$, $A_{i}$, $H_{i}^{\pm}$, 
and $H^{\pm \pm}$,
are given in terms of the mixing matrices $O$ as
\begin{align}
H_{i} = (O_{R})_{ia} R_{a},
\quad
A_{i} = (O_{I})_{ia} I_{a},
\quad
H^{\pm}_{i} = (O_{\pm})_{ia} S^{\pm}_{a},
\quad
H^{\pm \pm} = \Delta^{\pm \pm}. 
\end{align}
The matrices $O_{I}$ and $O_{\pm}$ can be explicitly expressed in terms of the vevs as 
\begin{align}
(O_{I})_{ia}
=
\begin{pmatrix}
c v_{1} V^{2} & - 2 c v_{2} v_{3}^{2} & - c v_{2}^{2} v_{3}
\\
0 & \frac{v_{2}}{V} & -2 \frac{v_{3}}{V}
\\
\frac{1}{2} b \frac{v_{2}}{v_{1}}
&
b
&
\frac{1}{2} b \frac{v_{2}}{v_{3}}
\end{pmatrix},
\quad
(O_{\pm})_{ia}
=
\begin{pmatrix}
c_{\pm} & s_{\pm}
\\
-s_{\pm} & c_{\pm}
\end{pmatrix},
\label{eq:OI-Opm}
\end{align}
where
\begin{gather}
\label{eq:c-vevs}
V \equiv \sqrt{v_{2}^{2} + 4 v_{3}^{2}},
\quad
c \equiv
\frac{1}{\sqrt{v_{1}^{2} V^{4} + 4 v_{2}^{2} v_{3}^{4} + v_{2}^{4} v_{3}^{2}}},
\quad
b \equiv \frac{2 v_{1} v_{3}}{\sqrt{v_{1}^{2} V^{2} + v_{2}^{2} v_{3}^{2}}},
\\
c_{\pm} \equiv
\frac{v_{2}}{\sqrt{v_{2}^{2} + 2 v_{3}^{2}}},
\quad
s_{\pm} \equiv
\frac{\sqrt{2} v_{3}}{\sqrt{v_{2}^{2} + 2 v_{3}^{2}}}.
\end{gather}
In the mass eigenbasis specified by the mixing matrices in Eq.~\eqref{eq:OI-Opm}, 
the Nambu-Goldstone bosons and the Higgs bosons 
are identified as follows: 
\begin{align}
J = A_{1}, \quad G^{0} = A_{2}, \quad A = A_{3},
\quad G^{\pm} = H^{\pm}_{1}, \quad H^{\pm} = H^{\pm}_{2},
\label{eq:masseig}
\end{align}
where $J$ is the majoron field,
$G^{0}$ and $G^{\pm}$ become
the longitudinal components of the weak gauge bosons,
and $A$ and $H^{\pm}$ are CP-odd and charged Higgs bosons with mass
\begin{align}
M_{A}^{2} = \frac{1}{2} \kappa 
\left(
\frac{v_{1} v_{2}^{2}}{v_{3}}
+
\frac{v_{2}^{2} v_{3}}{v_{1}}
+
4 v_{1} v_{3}
\right),\quad
M_{H^{\pm}}^{2}
=
\frac{1}{2}
\left(
\kappa \frac{v_{1}}{v_{3}}
-
\frac{1}{2} \lambda_{5}
\right)
\left(
v_{2}^{2} + 2 v_{3}^{2}
\right).
\end{align}
Since the rank of the mass matrix $M_{R}^{2}$ is three in general, 
there are three CP-even Higgs bosons.
We are working in the small-mixing regime 
($\beta_{2,3} \ll 1$ 
and $\kappa \ll 1$), 
$(O_{R})_{ia} \simeq \delta_{ia}$,\footnote{%
Notice that, in Sec.~IV of Ref.~\cite{Bonilla:2015jdf}, the authors sort the masses 
and use the label $i=1,2,3$ from smallest to largest, while
here the label $H_{i=1,2,3}$ indicates the dominant component 
among $R_{a=1,2,3}$.
In short, $H_{1}$ is not the lightest neutral CP-even Higgs boson but the one that consists dominantly of the singlet scalar, which is the heaviest.}
and therefore, $H_{1}$ mainly consists of the singlet field, with mass $M_{H_{1}} \simeq 2 \beta_{1} v_{1}^{2}$, and is decoupled from the other scalars.
In addition, $H_{1}$ does not participate in any gauge interaction,
hence it remains out of equilibrium during the whole history of the universe.
The 125~GeV Higgs boson is identified to $H_{2}$, while $H_{3}$ is an extra CP-even Higgs boson, which has 
a mass of $M_{H_{3}}^{2} \simeq \frac{1}{2} \kappa \frac{v_{1} v_{2}^{2}}{v_{3}}$.

Since the extra Higgs bosons except for $H_{1}$, 
\emph{i.e.},~$H_{3}$, $A$, $H^{\pm}$, and $H^{\pm \pm}$,
are made dominantly of the triplet field $\Delta$, and 
their masses stem mainly from a common origin, 
which is the $\kappa$ term in Eq.~\eqref{eq:typeII-scalar-potential},
their masses are of a similar size:
\begin{align}
M_{H_{3}}^{2}
\simeq\,&
M_{A}^{2}
\simeq 
M_{\Delta}^{2}
\equiv
\frac{1}{2}
\kappa
\frac{v_{1} v_{2}^{2}}{v_{3}},
\\
M_{H^{\pm}}^{2}
\simeq\,&
M_{\Delta}^{2}
-
\frac{1}{4} \lambda_{5} v_{2}^{2},
\\
M_{H^{\pm \pm}}^{2}
\simeq\,&
M_{\Delta}^{2}
-
\frac{1}{2}
\lambda_{5} v_{2}^{2}.
\end{align}
The gauge interactions of the scalars are listed in the appendix of Ref.~\cite{Bonilla:2015jdf}.
We will give the necessary scalar cubic and quartic interactions.

The Yukawa interactions in this model are given as 
\begin{align}
\mathscr{L}_{Y}
=&
(Y_{\Delta})_{\alpha \beta} \overline{L^{c}}_{\alpha} \Delta L_{\beta}
+
(Y_{\ell})_{\alpha \beta} \overline{L}_{\alpha} \ell_{R \beta} \text{i}\sigma_{2} \Phi^{*}
\nonumber
\\
&+
(Y_{u})_{\alpha \beta} \overline{Q}_{\alpha} u_{R\beta} \Phi
+
(Y_{d})_{\alpha \beta} \overline{Q}_{\alpha} d_{R \beta} \text{i} \sigma_{2} \Phi^{*}
+
\text{H.c.}\, ,
\label{eq:L_Yukawa-typeII}
\end{align}
where 
$\ell_{R}$ is a right-handed lepton singlet,
$Q$ is a quark doublet,
$u_{R}$ and $d_{R}$ 
are up- and down-type quark singlets,
respectively,
and
${\rm i} \sigma_{2}$ is an anti-symmetric tensor for the indices of $SU(2)_{L}$ fundamental representations.
This provides the fermion masses and
the interactions among the scalars and the fermions, 
\begin{align}
\mathscr{L}_{Y}
=&
-
\frac{1}{2} m_{\nu_{i}} \overline{{\nu_{L}}^{c}}_{i} \nu_{L i}
-
m_{\ell_{\alpha}}
\overline{\ell_{L}}_{\alpha} \ell_{R \alpha}
-
m_{u_{i}} \overline{u_{L}}_{i} u_{Ri}
-
m_{d_{i}} \overline{d_{L}}_{i} d_{Ri}
\nonumber
\\
&
-
\frac{m_{\nu_{i}}}{v_{3}}
(U_{\text{PMNS}}^{\dagger})_{i \alpha}
(O_{\pm})_{12} 
\overline{{\nu_{L}}^{c}}_{i} \ell_{L \alpha} G^{+}
-
\frac{m_{\nu_{i}}}{v_{3}}
(U_{\text{PMNS}}^{\dagger})_{i \alpha}
(O_{\pm})_{22} 
\overline{{\nu_{L}}^{c}}_{i} \ell_{L \alpha} H^{+}
\nonumber
\\
&-
\frac{1}{\sqrt{2}}(U_{\text{PMNS}}^{*})_{\alpha i} \frac{m_{\nu_{i}}}{v_{3}} (U_{\text{PMNS}}^{\dagger})_{i \beta}
\overline{{\ell_{L}}^{c}}_{\alpha} \ell_{L \beta} \Delta^{++}
\nonumber
\\
&
+
\frac{\sqrt{2} m_{\ell_{\alpha}}}{v_{2}} 
(O_{\pm})_{11} \overline{\nu_{L}}_{\alpha} \ell_{\alpha} G^{+}
+
\frac{\sqrt{2} m_{\ell_{\alpha}}}{v_{2}} 
(O_{\pm})_{21} \overline{\nu_{L}}_{\alpha} \ell_{\alpha} H^{+}
\nonumber
\\
&
- 
(V_{\text{CKM}}^{\dagger})_{ij} \frac{\sqrt{2} m_{u_{j}}}{v_{2}}
(O_{\pm})_{11}
\overline{d_{L}}_{i} u_{R j} G^{-}
- 
(V_{\text{CKM}}^{\dagger})_{ij} \frac{\sqrt{2} m_{u_{j}}}{v_{2}}
(O_{\pm})_{21}
\overline{d_{L}}_{i} u_{R j} H^{-}
\nonumber
\\
&
+
(V_{\text{CKM}})_{ij}
\frac{\sqrt{2} m_{d_{j}}}{v_{2}}
(O_{\pm})_{11}
\overline{u_{L}}_{i} d_{R j}
G^{+}
+
(V_{\text{CKM}})_{ij}
\frac{\sqrt{2} m_{d_{j}}}{v_{2}}
(O_{\pm})_{21}
\overline{u_{L}}_{i} d_{R j}
H^{+}
\nonumber
\\
&
-
\frac{1}{2} \frac{m_{\nu_{i}}}{v_{3}} (O_{R})_{j 3} \overline{{\nu_{L}}^{c}}_{i} \nu_{L i} H_{j}
-
\text{i}
\frac{1}{2} \frac{m_{\nu_{i}}}{v_{3}} (O_{I})_{j 3} \overline{{\nu_{L}}^{c}}_{i} \nu_{L i} A_{j}
\nonumber
\\
&
-
\frac{m_{\ell_{\alpha}}}{v_{2}}
(O_{R})_{i2}
\overline{\ell_{L}}_{\alpha} \ell_{R \alpha} H_{i}
+
\text{i} 
\frac{m_{\ell_{\alpha}}}{v_{2}}
(O_{I})_{i2}
\overline{\ell_{L}}_{\alpha} \ell_{R\alpha} A_{i}
\nonumber
\\
&
-
\frac{m_{u_{i}}}{v_{2}}
(O_{R})_{j2}
\overline{u_{L}}_{i}
u_{Ri}
H_{j}
-
\text{i}
\frac{m_{u_{i}}}{v_{2}}
(O_{I})_{j2}
\overline{u_{L}}_{i}
u_{Ri}
A_{j}
\nonumber
\\
&
-
\frac{m_{d_{i}}}{v_{2}}
(O_{R})_{j2}
\overline{d_{L}}_{i}
d_{Ri}
H_{j}
+
\text{i}
\frac{m_{d_{i}}}{v_{2}}
(O_{I})_{j2}
\overline{d_{L}}_{i}
d_{Ri}
A_{j}
+
\text{H.c.}\,,
\label{eq:L-Yukawa-Nr2}
\end{align}
where the masses of the fermions are given by 
\begin{gather}
m_{\nu_{i}}
\equiv - \sqrt{2} 
v_{3}
\left(
Y_{\Delta}^{\text{diag}} 
\right)_{i},
\quad
m_{\ell_{\alpha}}
\equiv
\frac{v_{2}}{\sqrt{2}}
\left(Y_{\ell}^{\text{diag}} \right)_{\alpha},
\\
m_{u_{i}} 
\equiv
-
\frac{v_{2}}{\sqrt{2}}
\left(Y_{u}^{\text{diag}} \right)_{i},
\quad
m_{d_{i}} 
\equiv
\frac{v_{2}}{\sqrt{2}}
\left( Y_{d}^{\text{diag}} \right)_{i},
\end{gather}
in terms of the diagonal matrices $Y_X^{\text{diag}}$, obtained by diagonalising the Yukawa matrices in Eq.~(\ref{eq:L_Yukawa-typeII}).

Although the anomaly contribution to the $J\gamma \gamma$ interaction cancels, the majoron in this model has an interaction with photons through
(i) majoron-pion mixing and (ii) triangle diagrams that depends on the mass of the fermions 
in the loop.
The contribution (i) is calculated 
in Sec.~7 of Ref.~\cite{Bauer:2020jbp}.
The coefficients in the formula Eq.~(92) in
Ref.~\cite{Bauer:2020jbp}
correspond to the parameters of the type-II majoron model
as
\begin{align}
\frac{c_{uu}}{f} = -2 cv_{3}^{2},
\quad
\frac{c_{dd}}{f} = 2 cv_{3}^{2},
\quad 
c_{\gamma\gamma}
=
c_{GG} = 0.
\label{eq:c_qq-typeII}
\end{align}
where $c$ is the combination of vevs given in \eref{eq:c-vevs}.
The contribution (ii) is discussed in Sec.~4 of Ref.~\cite{Bauer:2020jbp}.
For majorons with a mass in the range that we are interested in, only the contributions from first-generation fermions are relevant.
The coefficients for the up- and down-type quarks are given above.
For the electron-loop contribution, we need the coefficient $c_{ee}$, which is given as
\begin{align}
\frac{c_{ee}}{f}
= 2 c v_{3}^{2}.
\label{eq:c_ee-typeII}
\end{align}
Substituting Eqs.~\eqref{eq:c_qq-typeII} and \eqref{eq:c_ee-typeII} into Eqs.~(46) and (47) of Ref.~\cite{Bauer:2020jbp}, we can obtain the loop contributions in the majoron-photon coupling $g_{J \gamma \gamma}$.\footnote{%
With the majoron interactions in Lagrangian Eq.~\eqref{eq:L-Yukawa-Nr2}, we obtain $B_{1}(\tau_{f}) - 1$ instead of $B_{1}(\tau_{f})$ in Eq.~(47) of Ref.~\cite{Bauer:2020jbp} as a contribution from the loop of the fermion $f$~\cite{Bauer:2017ris}.
However, the ``$-1$'' part cancels summing over the contributions from different fermions in each generation, because the lepton number is anomaly free.
In the final expression Eq.~\eqref{eq:g_Jphph-typeII-in-app},
we only have $B_{1}(\tau_{f})$ as in Ref.~\cite{Bauer:2020jbp}, which correctly decouples in the heavy fermion limit, $m_f \gg M_J$.}

We can now summarise the majoron couplings in the type-II majoron model:
\begin{align}
\mathscr{L}_{J}
=
{\rm i}
g_{Jff}^{P}\, J\,
\overline{f} \gamma^{5} f
-
\frac{1}{4}
g_{J \gamma \gamma}\, J\,
F_{\mu \nu} \widetilde{F}^{\mu \nu},
\end{align}
with 
\begin{gather}
g_{J \nu_{i} \nu_{i}}^{P}
=
-
\frac{1}{2} m_{\nu_{i}} c v_{2}^{2},
\quad
g_{J \ell_{\alpha} \ell_{\alpha}}^{P}
=
-
2 m_{\ell_{\alpha}} c v_{3}^{2},
\quad
g_{J u_{i} u_{i}}^{P}
=
2 m_{u_{i}} c v_{3}^{2},
\quad
g_{J d_{i} d_{i}}^{P}
=
-
2 m_{d_{i}} c v_{3}^{2},
\\
g_{J\gamma \gamma}
=
\frac{2\alpha}{\pi}
cv_{3}^{2}
\left[ 
\frac{M_{J}^{2}}{M_{J}^{2} - m_{\pi^{0}}^{2}}
-
B_{1}(\tau_{e})
+
\frac{4}{3} B_{1}(\tau_{u})
-
\frac{1}{3} B_{1}(\tau_{d})
\right],
\label{eq:g_Jphph-typeII-in-app}
\end{gather}
where the quantity $c$ appearing in the above expressions, for the regime of parameters as in~\eref{eq:vev-hie}, is approximately given by $c \simeq (v_1 v_2^2)^{-1}$, and the loop function $B_{1}$ reads~\cite{Bauer:2017ris}
\begin{align}
B_{1}(\tau_{f})
\equiv\,&
1 - \tau_{f} f^{2} (\tau_{f}),
\qquad
f(\tau_{f}) \equiv  
\begin{cases}
\arcsin \frac{1}{\sqrt{\tau_{f}}},  &\tau_{f} \geq 1,
\\
\frac{\pi}{2} + \frac{\rm i}{2} \ln \frac{1 + \sqrt{1-\tau_{f}}}{1-\sqrt{1-\tau_{f}}}, \quad&\tau_{f} <1,
\end{cases}
\end{align}
and 
$\tau_{f} \equiv 4 m_{f}^{2}/ M_{J}^{2}$.
 Notice that above approximate expression for the majoron-pion mixing term is valid in the small mixing regime, $|m_\pi^2 - M_J^2| \gg (f_\pi/ v_1) m_J  m_{\pi^0}$ and $M_J\neq m_{\pi^0}$~\cite{Bauer:2017ris}.
The resulting decay rates relevant for a majoron with a mass in the keV-MeV range are given in  Eqs.~(\ref{eq:Jgammas})-(\ref{eq:Jee}).

\section{Relic density of type-II majorons from freeze-in}
\label{App:RelicDensity}

The decay rates relevant for the freeze-in production of type-II majorons are\,\footnote{%
Depending on the mass hierarchy of the triplet scalars,
$H_{3} \rightarrow A J$ or $A \rightarrow H_{3} J$
also produce majorons. However, the 
masses are almost degenerate and the decay rates are
suppressed in comparison to the main production processes in
Eq.~\eqref{eq:main-freezein}.}
\begin{align}
 \Gamma(H^{\pm} \rightarrow W^{\pm} J)=&
 \frac{g^{2} \left|
 s_{\pm} (O_{I})_{12} + \sqrt{2} c_{\pm} (O_{I})_{13}
 \right|^{2}
 }{64 \pi }
 \frac{M_{H^{\pm}}^{3}}{M_{W}^{2}}
 \left[
 1
 -
 \frac{M_{W}^{2}}{M_{H^{\pm}}^{2}}
 \right]^{3},
 \label{eq:Rate-Hpm-W-J}
 \\
 \Gamma(H_{3} \rightarrow Z J)=&
 \frac{
 g^{2}
 \left|
 (O_{R})_{32} (O_{I})_{12} - 2 (O_{R})_{33} (O_{I})_{13}
 \right|^{2}
 }{64 \pi c_{W}^{2}}
 \frac{M_{H_{3}}^{3}}{M_{Z}^{2}}
 \left[
 1 - \frac{M_{Z}^{2}}{M_{H_{3}}^{2}}
 \right]^{3},
 \\
\Gamma(A \rightarrow H_{2} J)=&
 \frac{|g_{H_{2} A J}|^{2}}{16 \pi M_{A}} 
 \left[
 1 - \frac{M_{H_{2}}^{2}}{M_{A}^{2}}
 \right],
 \label{eq:Rate-A-HSM-J} 
 \\ 
\Gamma(H_{3} \rightarrow J J)=&
 \frac{1}{32 \pi}
 M_{H_{3}}^{3}
 \left[
 \frac{((O_{I})_{11})^{2}}{v_{1}}
 (O_{R})_{31}
 +
 \frac{((O_{I})_{12})^{2}}{v_{2}}
 (O_{R})_{32}
 +
 \frac{((O_{I})_{13})^{2}}{v_{3}}
 (O_{R})_{33}
 \right]^{2},
\end{align}
where the cubic coupling $g_{H_{i} A J}$ is given as
\begin{align}
g_{H_{i} A J}
=&
 \Bigl[
 \left(
 \beta_{2}
 v_{1}
 +
 \kappa v_{3}
 \right)
 (O_{R})_{i1} 
 +
 2
 \lambda_{1}
 v_{2}
 (O_{R})_{i2}
 +
 \left(
 \left( \lambda_{3} + \lambda_{5} \right)
 v_{3}
 +
 \kappa v_{1}
 \right)
 (O_{R})_{i3} 
 \Bigr]
 (O_{I})_{12}
 (O_{I})_{32}
 \nonumber
 \\
 &
 +
 \Bigl[
 \beta_{3}
 v_{1}
 (O_{R})_{i1} 
 +
 \left(\lambda_{3} + \lambda_{5} \right)
 v_{2}
 (O_{R})_{i2}
 +
 2
 \left(\lambda_{2}+\lambda_{4}\right)
 v_{3}
 (O_{R})_{i3} 
 \Bigr]
 (O_{I})_{13} 
 (O_{I})_{33}
 \nonumber
 \\
 &
 +
 \Bigl[
 2
 \beta_{1}
 v_{1}
 (O_{R})_{i1} 
 +
 \beta_{2}
 v_{2}
 (O_{R})_{i2} 
 +
 \beta_{3}
 v_{3}
 (O_{R})_{i3} 
 \Bigr]
 (O_{I})_{11} 
 (O_{I})_{31}  
 \nonumber
 \\
 &
 +
 \kappa v_{2}
 (O_{R})_{i2} 
 \Bigl[
 (O_{I})_{11} 
 (O_{I})_{33}  
 +
 (O_{I})_{13} 
 (O_{I})_{31}  
 \Bigr]
 \nonumber
 \\
 &
 +
 \kappa 
 \Bigl[
 v_{2}
 (O_{R})_{i1} 
 +
 v_{1}
 (O_{R})_{i2}
 \Bigr]
 \Bigl[
 (O_{I})_{12} 
 (O_{I})_{33}  
 +
 (O_{I})_{13} 
 (O_{I})_{32}  
 \Bigr]
 \nonumber
 \\
 &
 +
 \kappa
 \Bigl[
 v_{2}
 (O_{R})_{i3}
 +
 v_{3}
 (O_{R})_{i2} 
 \Bigr]
 \Bigl[
 (O_{I})_{11} 
 (O_{I})_{32}  
 +
 (O_{I})_{12} 
 (O_{I})_{31}  
 \Bigr].
\end{align}
In the parameter region we are interested in,
the decay rates of Eqs.~\eqref{eq:Rate-Hpm-W-J}-\eqref{eq:Rate-A-HSM-J} 
can be approximated
to
\begin{align}
\sum_{\pm} \Gamma(H^{\pm} \rightarrow W^{\pm} J )
\simeq
\Gamma(H_{3} \rightarrow Z J)
\simeq
\Gamma(A \rightarrow H_{2} J)
\simeq
\frac{\kappa^{2} v_{2}^{2}}{16 \pi M_{\Delta}},
\label{eq:main-freezein}
\end{align}
which are the main engine of the freeze-in production of majorons.
The decay rate of $H_{3} \rightarrow JJ$ is significantly smaller than 
them.
The SM Higgs boson also decays into majoron(s), and
the decay rates are
\begin{align}
\Gamma(H_{2} \rightarrow Z J)=&
\frac{
 g^{2}
 \left|
 (O_{R})_{22} (O_{I})_{12} - 2 (O_{R})_{23} (O_{I})_{13}
 \right|^{2}
 }{64 \pi c_{W}^{2}}
 \frac{M_{H_{2}}^{3}}{M_{Z}^{2}}
 \left[
 1 - \frac{M_{Z}^{2}}{M_{H_{2}}^{2}}
 \right]^{3},
 \\
 \Gamma(H_{2} \rightarrow J J)=&
 \frac{1}{32 \pi}
 M_{H_{2}}^{3}
 \left[
 \frac{((O_{I})_{11})^{2}}{v_{1}}
 (O_{R})_{21}
 +
 \frac{((O_{I})_{12})^{2}}{v_{2}}
 (O_{R})_{22}
 +
 \frac{((O_{I})_{13})^{2}}{v_{3}}
 (O_{R})_{23}
 \right]^{2}.
\end{align}
They only contribute to the majoron production sub-dominantly.

The relic density of majorons produced through $B_{1} \rightarrow B_{2} J$
and $B \rightarrow J J$ results~\cite{Hall:2009bx}
\begin{align}
\Omega_{J} h^{2}
\simeq&
\frac{1.09 \cdot 10^{27}}{g_{*s}(M_{B_{1}}) \sqrt{g_{*} (M_{B_{1}}) } }
\frac{\Gamma(B_{1} \rightarrow B_{2} J) M_{J}}{M_{B_{1}}^{2}},
\label{eq:OmegahSq-FI-BBJ}
\\
\Omega_{J} h^{2}
\simeq&
\frac{2.18 \cdot 10^{27}}{g_{*s}(M_{B}) \sqrt{g_{*} (M_{B}) } }
\frac{\Gamma(B \rightarrow J J) M_{J}}{M_{B}^{2}},
\end{align}
where $g_{*}(T)$ is the relativistic degrees of freedom
at a temperature $T$,
and $g_{*s}(T)$ is the effective degrees of freedom for entropy
at $T$.

Following the procedure developed in Sec.~6.3 in Ref.~\cite{Hall:2009bx}, 
we can calculate the relic density of majoron produced from 
$2\rightarrow 2$ scattering processes with
the scalars $\Delta$, $\Phi_{1}$, $\Phi_{2}$, and $J$,
where $\Delta$ and $\Phi_{i}$ denote the components of triplet and doublet scalar fields, respectively. 
Here we have to take into account the masses of the bath particles, which are neglected in the calculation in Ref.~\cite{Hall:2009bx}.
Setting the mass $M_{\Delta}$ of the heaviest particle in the process as the lower limit of the momentum in the integration, the Boltzmann equation reads
\begin{align}
\frac{{\rm d} n_{J}}{{\rm d} t}
+
3 H n_{J}
\simeq&
\frac{2T}{512 \pi^{6}}
\int_{M_{\Delta}^{2}}^{\infty}
{\rm d}s {\rm d} \Omega 
P_{\Delta \Phi} 
P_{\Phi J}
|\mathcal{M}|^{2}
\frac{K_{1} (\sqrt{s}/T )}{\sqrt{s}}
\nonumber
\\
&
+
\frac{T}{512 \pi^{6}}
\int_{M_{\Delta}^{2}}^{\infty}
{\rm d}s {\rm d} \Omega 
P_{\Phi \Phi} 
P_{\Delta J}
|\mathcal{M}|^{2}
\frac{K_{1} (\sqrt{s}/T )}{\sqrt{s}}
\nonumber
\\
\simeq&
\frac{3|\kappa|^{2}}{128 \pi^{5}}
M_{\Delta} 
T^{3}
K_{1} (M_{\Delta}/T),
\label{eq:Boltzmann-eq-FI-scattering}
\end{align}
where
$K_{1}(x)$ is the modified Bessel function of the first kind, and
$P_{ij}$ is a function introduced in Ref.~\cite{Hall:2009bx}, which is defined as
\begin{align}
P_{ij}
\equiv
\frac{
\sqrt{ s - (m_{i} + m_{j})^{2} } 
\sqrt{ s - (m_{i} - m_{j})^{2} }}
{2 \sqrt{s}},
\end{align}
with $s$ denoting the centre-of-mass energy.
Defining $Y_{J}(T) \equiv n_{J}(T)/s(T) $,
we can write the Boltzmann equation as
\begin{align}
Y_{J}(T_{\text{eq}})
=
\frac{3|\kappa|^{2}}{128 \pi^{5}}
M_{\Delta} 
\int_{T_{\text{eq}}}^{\infty}
{\rm d} T\,
T^{2}
\frac{K_{1} (M_{\Delta}/T)}{H(T) s(T)},
\end{align}
where $s(T)$ is the entropy density at $T$, \emph{not} the centre-of-mass
energy.
In the radiation-dominated era $T>T_{\text{eq}}$\,\,---\,\,
where $T_{\text{eq}}$ is the temperature of the epoch of
matter-radiation equality\,\,---\,\,the Hubble parameter is 
\begin{align}
H(T) \simeq & \sqrt{\frac{4\pi^{3}}{45} g_{*}(T)} \frac{1}{M_{\text{Pl}}} T^{2},
\end{align}
and the entropy density is expressed 
in terms of the effective degrees of freedom as
\begin{align}
s(T) = \frac{2\pi^{2}}{45} g_{*s} (T) T^{3}.
\end{align}
With $Y_{J}(T_{\text{eq}}) = Y_{J}(T_{0})$,
the relic density resulting from the scattering processes is\,\footnote{%
Note that Eq.~\eqref{eq:OmegaJhSq-scattering} 
includes the contributions from all three possible 
scattering 
processes, \emph{i.e.},~$\Phi_{1} \Phi_{2} \rightarrow \Delta J$,
$\Phi_{1} \Delta \rightarrow \Phi_{2} J$,
and 
$\Phi_{2} \Delta \rightarrow \Phi_{1} J$,
cf. Eq.~\eqref{eq:Boltzmann-eq-FI-scattering}}
\begin{align}
\Omega_{J} h^{2}
=
\frac{Y_{J} (T_{0}) s_{0}}{\rho_{c0}} M_{J} h^{2}
\simeq
\frac{5.55 \cdot 10^{23}}{g_{*s}(M_{\Delta}) 
\sqrt{g_{*} (M_{\Delta}) } }
|\kappa|^{2}
\frac{M_{J}}{M_{\Delta}},
\label{eq:OmegaJhSq-scattering}
\end{align}
where $s_{0}$, $\rho_{c0}$, and $T_{0}$ are 
the entropy, the critical density, and 
the temperature today.

The terms in the scalar potential, which contribute to the scattering processes are 
\begin{align}
- \mathscr{L} 
\supset& 
\kappa
\Biggl[ 
\frac{1}{2} I_{1} R_{2}^{2} I_{3}
-
\frac{1}{2} I_{1} I_{2}^{2} I_{3}
+
I_{1} R_{2} I_{2} R_{3}
-
{\rm i} 
\frac{1}{\sqrt{2}} 
I_{1} R_{2} \phi^{-} \Delta^{+}
+
{\rm i} 
\frac{1}{\sqrt{2}} 
I_{1} R_{2} \phi^{+} \Delta^{-}
\nonumber
\\
&\hspace{0.5cm}
+
\frac{1}{\sqrt{2}} 
I_{1} I_{2} \phi^{-} \Delta^{+}
+
\frac{1}{\sqrt{2}} 
I_{1} I_{2} \phi^{+} \Delta^{-}
-
{\rm i}
\frac{1}{\sqrt{2}} 
I_{1} \phi^{-} \phi^{-} \Delta^{+ +} 
+
{\rm i}
\frac{1}{\sqrt{2}} 
I_{1} \phi^{+} \phi^{+} \Delta^{- -} 
\Biggr].
\end{align}

Freeze-in DM production in an early matter-dominated era
is discussed in Ref.~\cite{Co:2015pka,Calibbi:2021fld}.
Following Sec.~B.2 in Ref.~\cite{Calibbi:2021fld}, 
here we derive the formula of the relic density of majoron 
produced from the decay process $B_{1} \rightarrow B_{2} J$.
Here we start with Eq.~(B.20) in Ref.~\cite{Calibbi:2021fld},
which is
the quantity $\mathcal{X}_{J} \equiv n_{J} a^{3}$
at the reheating temperature $T_{R}$,\footnote{%
Note that $M_{\text{Pl}}$ in Ref.~\cite{Calibbi:2021fld}
is the {\it reduced} Planck mass.}
\begin{align}
\mathcal{X}_{J}(a_{R}) 
\simeq 
\sqrt{\frac{3}{8 \pi }}
M_{\text{Pl}}
\int_{0}^{a_{R}}
{\rm d} a
\frac{a^{2}}{\sqrt{\rho_{M}(a) }}
\mathcal{C}_{B_{1} \rightarrow B_{2} J}(a),
\label{eq:X-MDE-decay}
\end{align}
where
$a_{R}$ is the scale factor at $T_{R}$
and 
$\rho_{M}(a)$ is the energy density of the matter component,
which are explicitly given in Appendix \ref{App:EarlyMD}.
The collision term 
$\mathcal{C}_{B_{1} \rightarrow B_{2} J}(a)$
of the process $B_{1} \rightarrow B_{2} J$ is calculated to be
\begin{align}
\mathcal{C}_{B_{1} \rightarrow B_{2} J}(a)
\simeq
\frac{M_{B_{1}}^{2} \Gamma(B_{1} \rightarrow B_{2} J)}{2\pi^{2}}
T K_{1}
\left(
\frac{M_{B_{1}}}{T}
\right),
\end{align}
and the temperature $T$ is related to the scale factor $a$ as
$T=T_{R} a_{R}^{3/8}/a^{3/8}$
in the early matter-dominated era,
cf. Eq.~\eqref{eq:T-a-in-EMD}.
Substituting this into Eq.~\eqref{eq:X-MDE-decay},
we have 
\begin{align}
\mathcal{X}_{J}
(a_{R})
\simeq 
\sqrt{\frac{90}{8\pi^{3}}}
\frac{M_{\text{Pl}}}{T_{R}}
\frac{M_{B_{1}}^{2} \Gamma(B_{1} \rightarrow B_{2} J)}{2\pi^{2}}
\frac{a_{R}^{3}}{\sqrt{g_{*} (a_{R})}}
\int_{0}^{1} 
{\rm d} x
x^{25/8} K_{1} \left(\frac{M_{B_{1}}}{T_{R}} x^{3/8} \right),
\end{align}
where $x \equiv a/a_{R}$.
With 
$
Y_{J} (T_{0})
= 
Y_{J} (a_{R})
=
\frac{\mathcal{X}(a_{R})}{a_{R}^{3} s (T_{R})}$,
we arrive at the relic density formula
\begin{align}
\Omega_{J} h^{2}
\simeq&
\frac{2.33 \cdot 10^{26}}{g_{*s}(T_{R}) \sqrt{g_{*} (T_{R}) } }
\left[
\int_{0}^{1} \text{d}x x^{25/8} K_{1} \left( \frac{M_{B_{1}}}{T_{R}} x^{3/8} \right)
\right]
\frac{\Gamma(B_{1} \rightarrow B_{2} J) M_{B_{1}}^{2} M_{J}}{T_{R}^{4}}.
\end{align}
This result reproduces the power-law of the formula presented in Ref.~\cite{Calibbi:2021fld}, Eq.~(2.11).

One can calculate the contribution from $2 \rightarrow 2$ scattering processes in the same way.
The result is
\begin{align}
\Omega_{J} h^{2}
\simeq&
\frac{3.53 \cdot 10^{23}}{g_{*s}(T_{R}) \sqrt{g_{*} (T_{R}) } }
\left[
\int_{0}^{1} \text{d}x x^{19/8} K_{1} \left( \frac{M_{\Delta}}{T_{R}} x^{3/8} \right)
\right]
|\kappa|^{2}
\frac{M_{\Delta} M_{J}}{T_{R}^{2}}.
\end{align}
In the range of $T_{R} \simeq 20-50$ GeV, 
the degrees of freedom are 
$g_{*}(T_{R}) = g_{*s}(T_{R}) \simeq 90$.

\section{Early matter-dominated era}
\label{App:EarlyMD}

The cosmology of an early matter-dominated era is
discussed in, \emph{e.g.},~Refs.~\cite{Co:2015pka,Calibbi:2021fld}.
Here we derive the equations and relations that we used in our discussion on the freeze-in production of majorons 
in the early matter-dominated era.

Suppose that a non-relativistic matter field 
dominates the energy budget of the universe
before the radiation-dominated era,
and the matter field gradually decays to radiation
and injects entropy into the universe\,\,---\,\,an example being provided by the oscillations of the inflaton field during the epoch of its decay.
Energy-momentum conservation 
${(T_{\text{total}})^{\mu \nu}}_{;\mu} = 0$
determines the time-evolution of the energy densities of 
the matter and radiation components, 
$\rho_{M}$ and $\rho_{R}$, as
\begin{align}
&\frac{{\rm d} \rho_{M}}{{\rm d} t} 
+
3 H \rho_{M} = - \Gamma_{M} \rho_{M},
\\
&\frac{{\rm d} \rho_{R}}{{\rm d} t} 
+
4 H \rho_{R} = \Gamma_{M} \rho_{M},
\label{eq:drho_R-dt-in-EMD}
\end{align}
where $\Gamma_{M}$ is the decay rate of the matter particle and $H\equiv \dot{a}/a$ is the Hubble parameter.
Since the decay of the matter field progresses gradually
and the evolution of $\rho_{M}$ is dominantly described  
by the dilution due to the expansion of the universe,
the matter component evolves as
\begin{align}
\rho_{M}(t)
=
\rho_{M \text{in}}
\left(
\frac{a_{\text{in}}}{a(t)}
\right)^{3},
\label{eq:rho_M-in-EMD}
\end{align}
where $\rho_{M \text{in}}$ and 
$a_{\text{in}}$ are 
the ``initial" values of the 
energy density of the matter component
and the scale factor,
\emph{i.e.},~the values at the time $t_{\text{in}}$ 
when the early matter-dominated
era began.
In the evolution of the radiation component, 
the source term from the decay of the matter field is 
important, and the solution of the differential equation 
Eq.~\eqref{eq:drho_R-dt-in-EMD} (with Eq.~\eqref{eq:rho_M-in-EMD})
is given as
\begin{align}
\rho_{R}(t)
=&
\rho_{R \text{in}}
\left(
\frac{a_{\text{in}}}{a(t)}
\right)^{4}
+
\frac{2}{5}
\sqrt{
\frac{3}{8\pi}
\rho_{M \text{in}}
} 
\Gamma_{M}
M_{\text{Pl}}
\left[ 
\left(
\frac{a_{\text{in}}}{a(t)}
\right)^{3/2}
-
\left(
\frac{a_{\text{in}}}{a(t)}
\right)^{4}
\right]
\nonumber
\\
\simeq&
\frac{2}{5}
\sqrt{
\frac{3}{8\pi}
\rho_{M \text{in}}
} 
\Gamma_{M}
M_{\text{Pl}}
\left(
\frac{a_{\text{in}}}{a(t)}
\right)^{3/2},
\label{eq:rho_R-in-EMD}
\end{align}
where 
$\rho_{R \text{in}}$ is the initial value of $\rho_{R}$.
Here we used the Friedmann equation
\begin{align}
H(a) = 
\sqrt{\frac{8\pi}{3}\rho_{\text{total}}} 
\frac{1}{M_{\text{Pl}}} 
\simeq  
\sqrt{\frac{8\pi}{3}\rho_{M}(a)} 
\frac{1}{M_{\text{Pl}}}
\label{eq:Hubble-EMD}
\end{align}
in the early matter-dominated era.

Since the radiation species are in thermal equilibrium and follow 
the statistical distribution, $\rho_{R} \propto T^{4}$\,\,---\,\,more concretely, $\rho_{R} = (\pi^{2}/30) g_{*}(T) T^{4}$\,\,---\,\,from Eq.~\eqref{eq:rho_R-in-EMD} we find
$\rho_{R} \propto a^{-3/2}$ in the early matter-dominated era.
Combining these two observations, 
we can find that the temperature $T$ 
is related to the scale factor $a$ as
\begin{align}
T = T_{R} \frac{a_{R}^{3/8}}{a^{3/8}},
\label{eq:T-a-in-EMD}
\end{align}
where $a_{R}$ and $T_{R}$ are the scale factor and 
the temperature when $\rho_{M} = \rho_{R}$.
We call $T_{R}$ reheating temperature.
With the relation Eq.~\eqref{eq:T-a-in-EMD},
we can rewrite the evolution of $\rho_{M}$ Eq.~\eqref{eq:rho_M-in-EMD} as a function of $T$;
The matter density is diluted due to the expansion 
of the universe as $\rho_{M} \propto 1/a^{3}$, which means 
$\rho_{M} \propto T^{8}$.
From the temperature-dependence of $\rho_{M}$ and $\rho_{R}$ 
and the definition of the reheating temperature,
we find 
\begin{align}
\rho_{R}(T) = \frac{\pi^{2}}{30} g_{*}(T) T^{4},
\quad
\rho_{M}(T) = \frac{\pi^{2}}{30} g_{*}(T) \frac{T^{8}}{T_{R}^{4}},
\label{eq:rho_RM-as-funcT-EMD}
\end{align}
in the early matter-dominated era.
Using the relation Eq.~\eqref{eq:T-a-in-EMD},
they can be rewritten as functions of $a$, which are used in 
our calculations of the relic density of majoron.

The scale factor $a_{R}$ can be given as a function of $T_{R}$
with the relation $T a =$const, 
which is valid in the radiation-dominated era.
The radiation-dominated era starts at $T_{R}$ and ends
at $T_{\text{eq}}$ which is the temperature of the 
matter-radiation equality era.
With $a_{\text{eq}} = 1/3400$ and $T_{\text{eq}} = 0.75$ eV,
we have
\begin{align}
a_{R}
=
\frac{0.75 \cdot 10^{-9} \text{GeV}}{3400} 
\frac{1}{T_{R}\text{[GeV]}},
\end{align}
which is used in the calculations of $\mathcal{X}_{J}$, \eref{eq:X-MDE-decay}.
Although one can solve the reheating temperature
in terms of $\Gamma_{M}$ as
\begin{align}
T_{R} \simeq 
\sqrt{\frac{2}{5}}
\left(
\frac{90}{8\pi^{3} g_{*} (T_{R})} 
\right)^{1/4}
\sqrt{\Gamma_{M} M_{\text{Pl}}},
\end{align}
it is convenient to give $T_{R}$ as a free parameter.

\bibliographystyle{JHEP}
\bibliography{TypeII-majoron}

\end{document}